\documentclass[a4paper,fleqn,usenatbib]{mnras}

\usepackage[T1]{fontenc}
\usepackage{ae,aecompl}

\usepackage{cellspace} %
\usepackage{graphicx}	
\usepackage{amsmath}	
\usepackage{amssymb}	
\usepackage{xfrac}
\usepackage{makecell}
\usepackage{multirow}
\usepackage{ulem}
\usepackage{bm}

\newcommand{\kms}{km\,s$^{-1}$}	

\title[High resolution ALMA observation of 49 Ceti]
      {High resolution ALMA observation of the $^{12}$CO(3-2) and 
350 GHz continuum emissions of the debris disc of 49 Ceti}
\author[P. T. Nhung et al.]{
{P. T. Nhung\thanks{E-mail: pttnhung@vnsc.org.vn}, D. T. Hoai, P. Tuan-Anh, P. N. Diep, N. T. Phuong, N. T. Thao}
\newauthor{ and P. Darriulat}
\\
Department of Astrophysics, Vietnam National Satellite Center, VAST, 18 Hoang Quoc Viet, Hanoi, Vietnam\\
}
\date{Accepted XXX. Received YYY; in original form ZZZ}

\pubyear{2017}

\begin{document}
\label{firstpage}
\pagerange{\pageref{firstpage}--\pageref{lastpage}}
\maketitle

\begin{abstract}

We present high resolution ALMA observations of the CO(3-2) and 350 GHz continuum emissions of the debris disc of 49 Ceti, known to be particularly rich in molecular gas in spite of its age. The main new results are: i) both CO and dust discs share a same position angle and a same inclination but the gas disc is more homogeneous, more central and thinner than the dust disc; ii) evidence is obtained for a significant deficit of observed CO(3-2) emission at Doppler velocities differing from the star systemic velocity by less than 1 \kms; iii) gas velocities are accurately measured and found Keplerian over a broad range of disc radii; iv) the observed CO(3-2) line width is dominated by Keplerian shear and upper limits are obtained to the intrinsic line width. Simple phenomenological models of both CO(3-2) and \mbox{350 GHz} continuum emissions are presented, requiring the use of only very few parameters. The results are discussed in the frame of currently favoured models.
\end{abstract}




\section{Introduction}
Recent observations of 49 Ceti have established that it is a bright debris disc, seen close to edge-on, surrounding a single A1V star at a distance of 59$\pm$1 pc from the Sun \citep{vanLeuwen2007}. The evidence for it being a debris disc is obtained from its age and from the properties of the dust. Its age, first evaluated from standard evolution models, is now obtained from the identification of the central star as a member of the Argus association, in the \mbox{40 Myr range} \citep{Zuckerman2012}. The disc is very dusty and its spectral energy distribution (SED), implying a dust mass of $\sim$0.3 Earth masses \citep{Wahhaj2007}, suggests the presence of two components forming a cold (50 to 100 K) outer and a warmer (150 to 200 K) inner discs (\citealt{Wahhaj2007, Hughes2008, Roberge2013}). Contrary to standard debris discs, which have lost their gas after a few Myr of age, 49 Ceti is seen to contain an important {amount} of carbon monoxide (CO), exceeding 2$\times$10$^{-4}$ Earth masses, from millimetre/sub-millimetre observations of the emission of molecular rotational states, CO(2-1) and CO(3-2). Early observations (\citealt{Zuckerman1995, Dent2005}) have been followed by higher resolution Sub-Millimetre Array (SMA) observations \citep{Hughes2008} giving evidence for a rotating ring with a central hole interpreted as resulting from UV dissociation from the central star. 49 Ceti shares the peculiarity of containing significant {amounts} of CO gas with four other debris discs {having an age equal or superior to 5 Myr:} $\beta$ Pictoris (\citealt{Fernandez2006, Dent2014}), HD 21997 (\citealt{Moor2011, Moor2013, Kospal2013}), HD 141569 \citep{White2016} and \mbox{HD 131835} \citep{Moor2015}. It has been suggested that its CO component, as well as part of its dust, are not primordial but the result of multiple collisions of small CO-rich comet-like objects \citep{Zuckerman2012}, reminiscent of the Kuiper Belt in the solar system, {and having} a total mass of the order of 400 Earth masses. Such an interpretation is qualitatively supported by recent observations of variable emission from fast grazing comets near the star \citep{Miles2016}. Far-UV observations along the line of sight pointing to the star \citep{Roberge2014} using the Hubble Space Telescope (HST) have given evidence for the presence of absorption lines from atomic species, in particular C and O, and for the absence of CO absorption lines. This is interpreted as the disc having an inclination $i$ with respect to the line of sight large enough for the star to be seen though the central hole of the CO disc, {while atomic species are still present much closer to the central star} (see also \citealt{Montgomery2012}).

\section{Observations and data reduction}
Emission from 49 Ceti was observed by ALMA on 14$^\textrm{th}$ November 2013 (project code: 2012.1.00195.S, PI: M. Hughes) during 39 minutes on both continuum (width of 15.016 MHz, from 342.994 to 358.010 GHz) and \mbox{$^{12}$CO(3-2)} line. It used 28 antennas with maximal baseline of \mbox{1284.3 m} and minimal baseline of 17.3 m. The data were reduced by the ALMA staff in 2015 and published on JVO portal (http://jvo.nao.ac.jp/portal). We have checked the quality of data reduction and found it unnecessary to make changes. The present article makes no attempt at producing a physical model of the emission but simply at proposing a model of the de-projected morphology and kinematics of the disc. Considerations such as our using clean map rather than dirty map quantities are therefore unimportant. The acceptance loss associated with the primary beam is only $\sim$5\% at a distance of 2 arcsec from the star but the effect of the lack of short baselines (zero-spacing problem) is much larger: the maximal recoverable scale is evaluated to be 6.2 arcsec. Combining both effects and approximating the loss of acceptance at large distances by a Gaussian having a full width at half maximum (FWHM) of 5.7 arcsec, we obtain crude estimates of the effect. In what follows, as most of the results of the present study are largely independent from the acceptance at large distances, we mention these estimated corrections only when sufficiently pertinent.

\begin{figure}
\centering
\includegraphics[height=4.5cm,trim=1.7cm 1.cm 1.5cm 2.cm,clip]{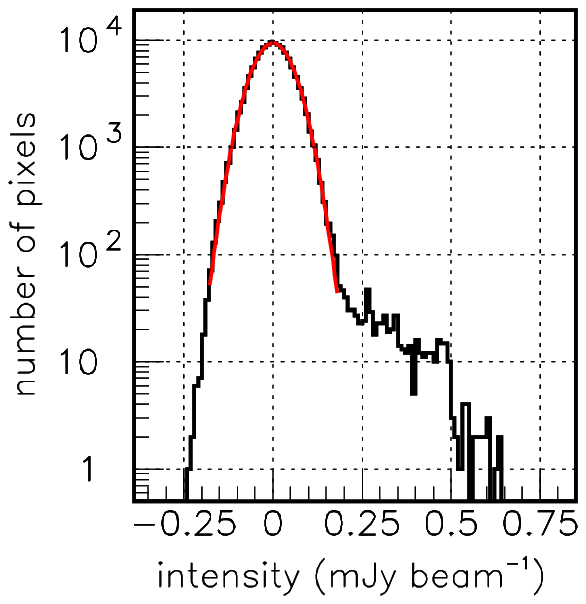}
\includegraphics[height=4.5cm,trim=2.cm 1.cm 2.3cm 2.cm,clip]{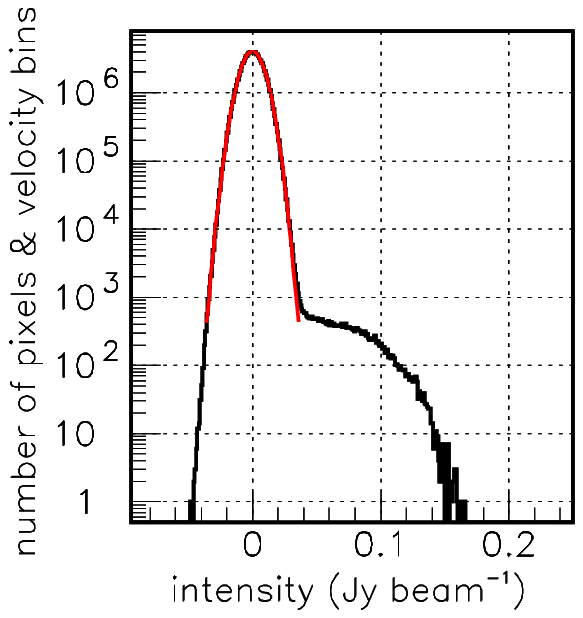}
\caption{Distribution of the continuum (left panel) and CO(3-2) (right panel) intensities.}
\label{fig1}
\end{figure}

The continuum data are presented in a grid of \mbox{360$\times$360 pixels}, each 0.1$\times$0.1 arcsec$^2$. The beam is 0.56$\times$0.43 arcsec$^2$ FWHM with a position angle of 88.6$^\circ$. Figure \ref{fig1} (left) displays the intensity distribution. A Gaussian fit of the noise gives a mean value of \mbox{5.6$\times$10$^{-5}$ mJy beam$^{-1}$} and an rms value of \mbox{0.055 mJy beam$^{-1}$}.

The $^{12}$CO(3-2) data are continuum subtracted and presented, as the continuum data, in a grid of 360$\times$360 pixels, each 0.1$\times$0.1 arcsec$^2$. The beam is 0.50$\times$0.38 arcsec$^2$ FWHM with a position angle of 83.7$^\circ$. Figure \ref{fig1} (right) displays the intensity distribution. A Gaussian fit of the noise gives a mean value of \mbox{$-$1.74$\times$10$^{-4}$ mJy beam$^{-1}$} and an rms value of \mbox{8.8 mJy beam$^{-1}$}. {Data cube elements are given in 650 bins of velocity, each 0.053 km s$^{-1}$ wide, covering between $-$14.6 and 19.8 km s$^{-1}$ with origin at 2.78 km s$^{-1}$ corresponding to the systemic velocity. The spectral resolution is evaluated as 0.1 km s$^{-1}$ (two velocity bins) by the ALMA staff.}

\section{Continuum and CO(3-2) emission: main features}
\subsection{Morphology}
\begin{figure}
\centering
\includegraphics[width=7.cm,trim=0.cm 8.cm 2.5cm 0.cm,clip]{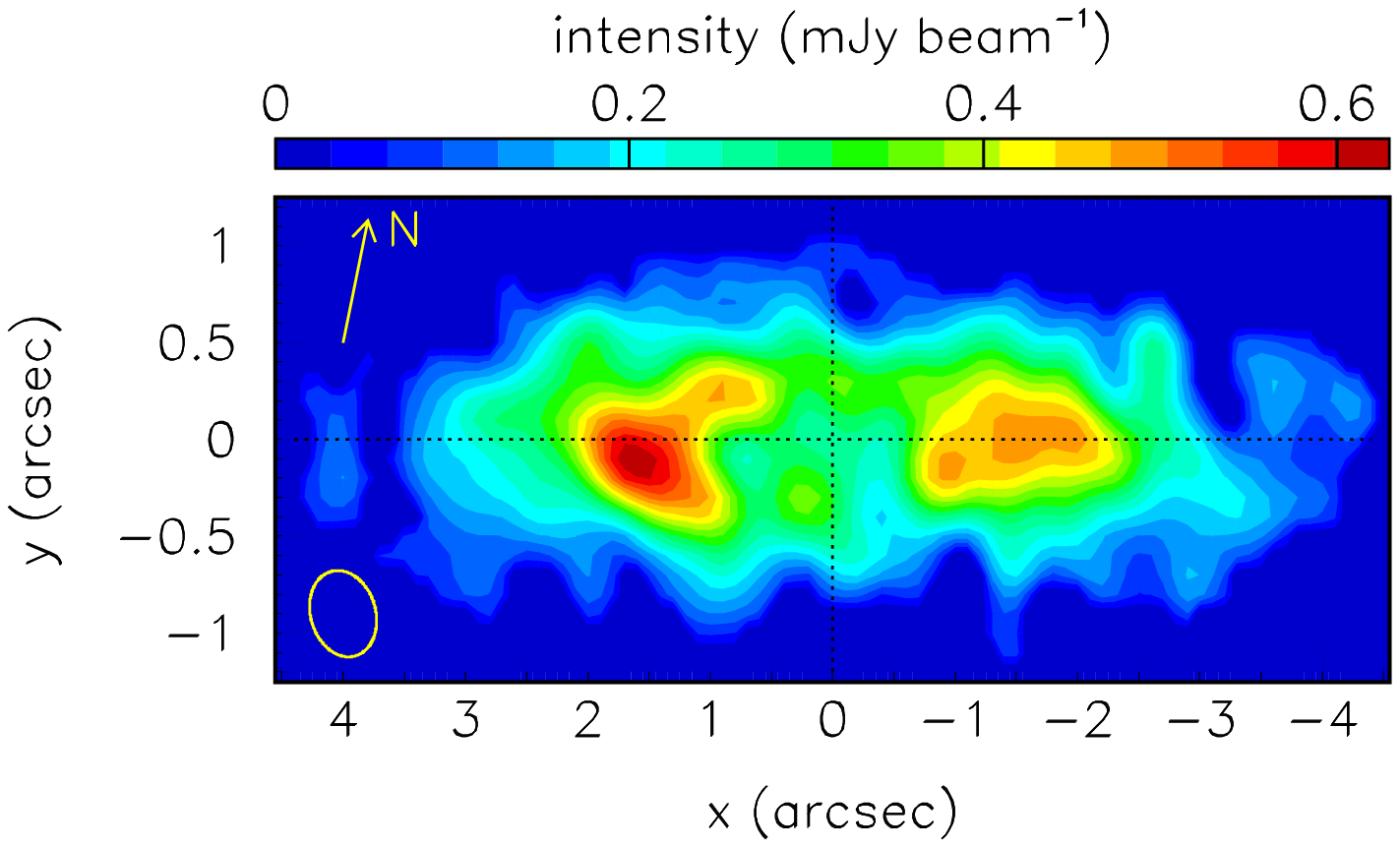}
\includegraphics[width=7.cm,trim=0.cm 8.cm 2.5cm 0.cm,clip]{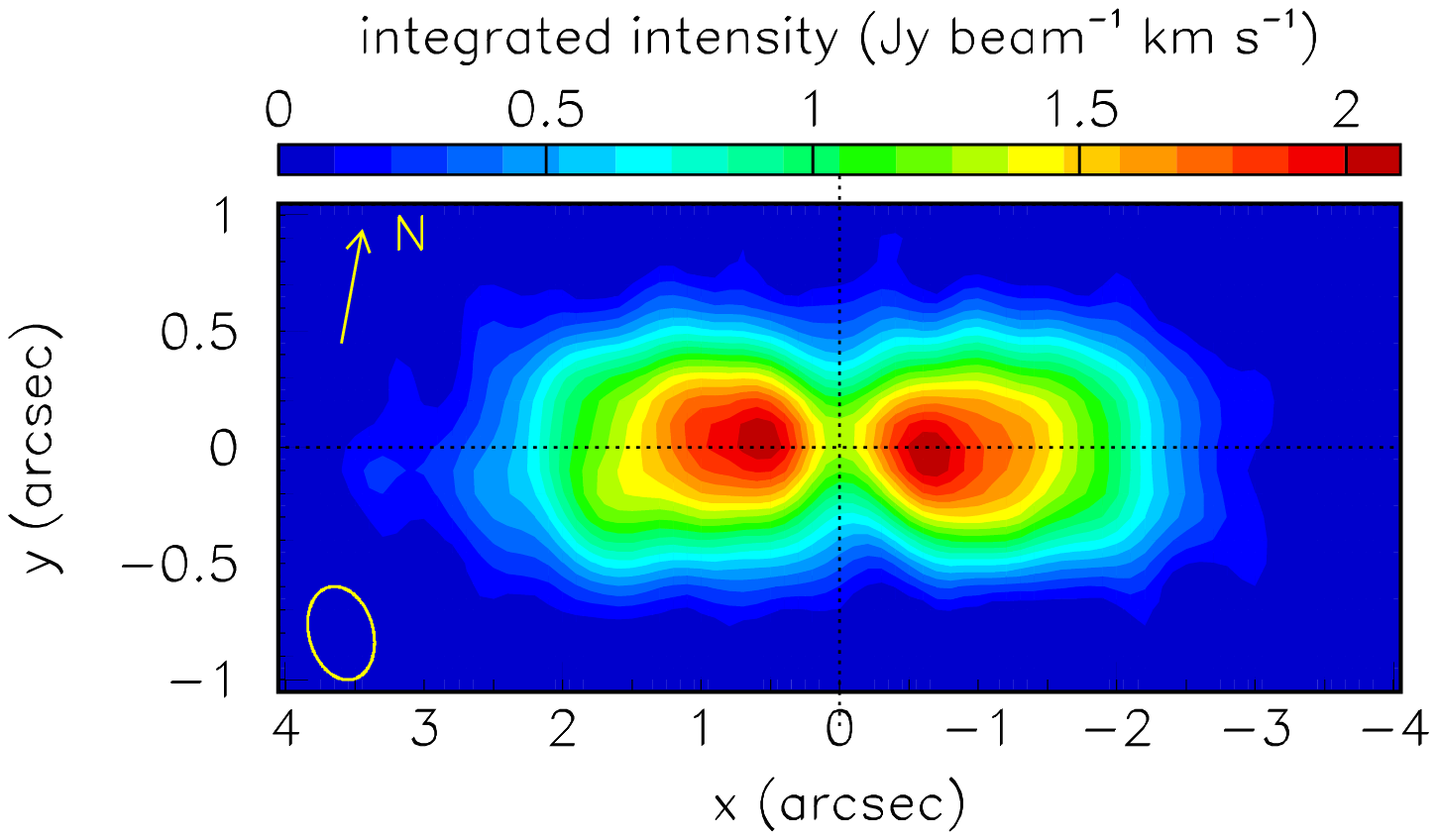}
\caption{Upper panel: cleaned sky map of the continuum emission (Jy beam$^{-1}$). Lower panel: cleaned sky map of the CO(3-2) emission \mbox{(Jy beam$^{-1}$ \kms)}. Note the different $x$ and $y$ scales. {The coordinate axes have been rotated as described in the text and shown by the arrow in the figures. The beams are shown in the lower left corners. One arcsec corresponds to 59 au.}}
\label{fig2}
\end{figure}

Figure \ref{fig2} displays the sky maps of the continuum emission (upper panel, 1-$\sigma$ cut) and of the velocity integrated \mbox{CO(3-2)} emission (lower panel) {limited to the central 8$\times$2 arcsec$^2$ region that covers well the disc emission}. In the CO case we use a 1-$\sigma$ cut, meaning 14.6 mJy beam$^{-1}$ \kms, on the velocity integrated intensity. We have checked that using stronger cuts, such as 2-{{$\sigma$}} and 3-$\sigma$, does not significantly affect any of the conclusions presented in the remainder of the article. For convenience, we use a frame of reference shifted west by 0.02 arcsec (where the dip in the gas intensity distribution peaks) and rotated clockwise by 17.5$^\circ$. {This value of the position angle was found to maximize the symmetry of the observed intensity with respect to the disc mid-plane when considering the line and continuum data jointly.} This frame of reference is used throughout the remainder of the article; the rotated west axis is labelled $x$, the rotated north axis is labelled $y$ and the $z$ axis points along the line of sight away from Earth.

\begin{figure}
\centering
\includegraphics[width=8.5cm,trim=1.cm 1.cm 4.cm 0.cm,clip]{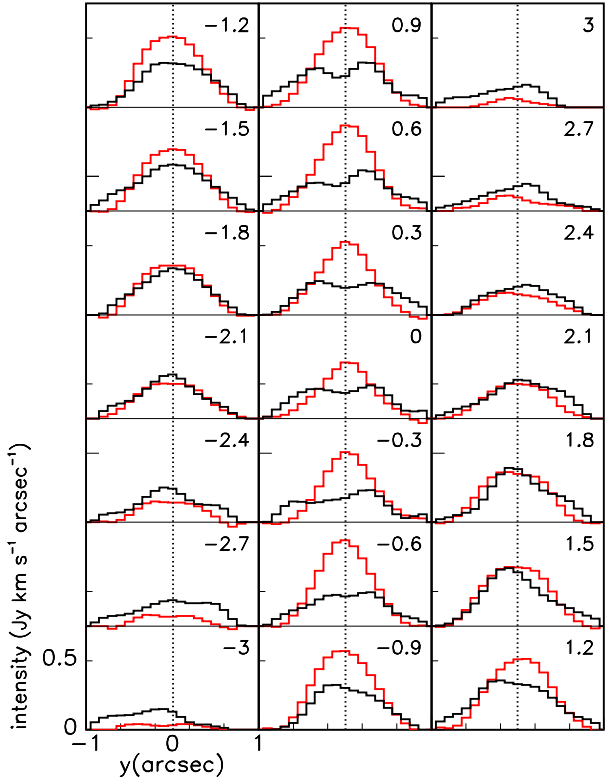}
\caption{Continuum (black) and CO(3-2) (red) emission: $y$-distribution of the intensity for various 0.3 arcsec wide intervals of $x$, in succession from the lower left corner to the upper right corner, starting at [$-$3.15 arcsec, $-$2.85 arcsec] and ending at [2.85 arcsec, 3.15 arcsec]. The central value of $x$ is inserted in each panel. {The curves have been normalized to a common area for convenience by scaling the continuum data up by a factor 620. }}
\label{fig3}
\end{figure}

\begin{figure}
\centering
\includegraphics[height=4.5cm,trim=1.7cm 1.cm 1.5cm 2.cm,clip]{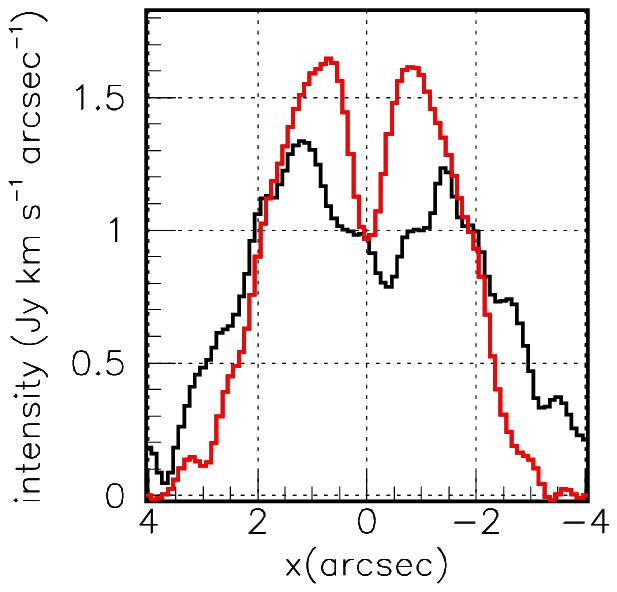}
\includegraphics[height=4.5cm,trim=2.cm 1.cm 2.3cm 2.cm,clip]{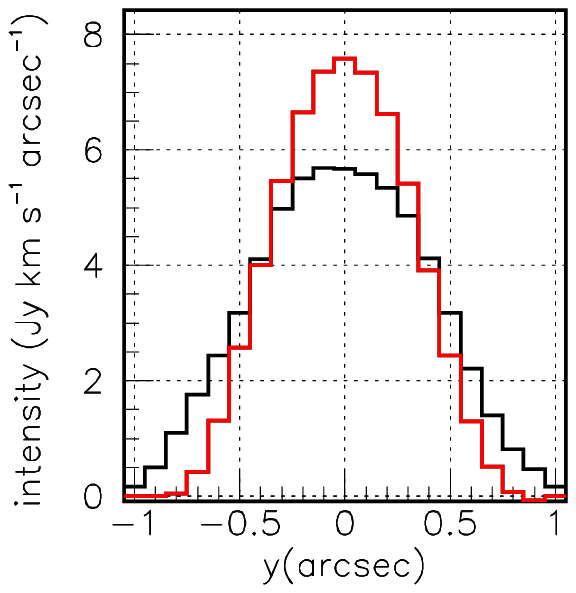}
\caption{Distribution of the observed intensity over $x$ (left) and over $y$ (right) for continuum (black) and CO(3-2) (red) emissions. The curves have been normalised to a common area for convenience by scaling the continuum data up by a factor 620. }
\label{fig4}
\end{figure}

\begin{figure*}
\centering
\includegraphics[height=5.5cm,trim=0.cm 1.cm 1.cm 2.cm,clip]{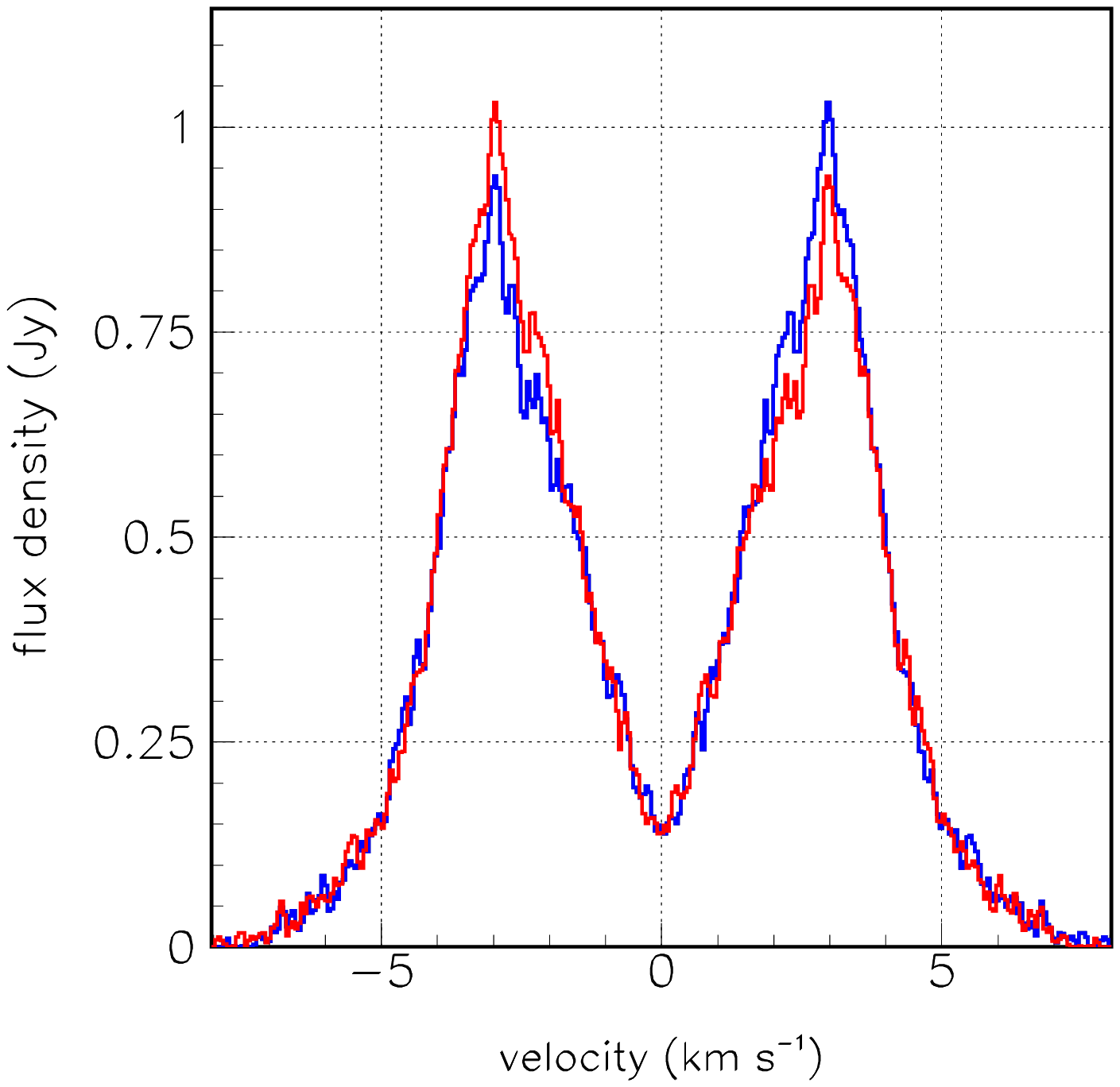}
\includegraphics[height=5.5cm,trim=0.cm 1.cm 1.cm 2.cm,clip]{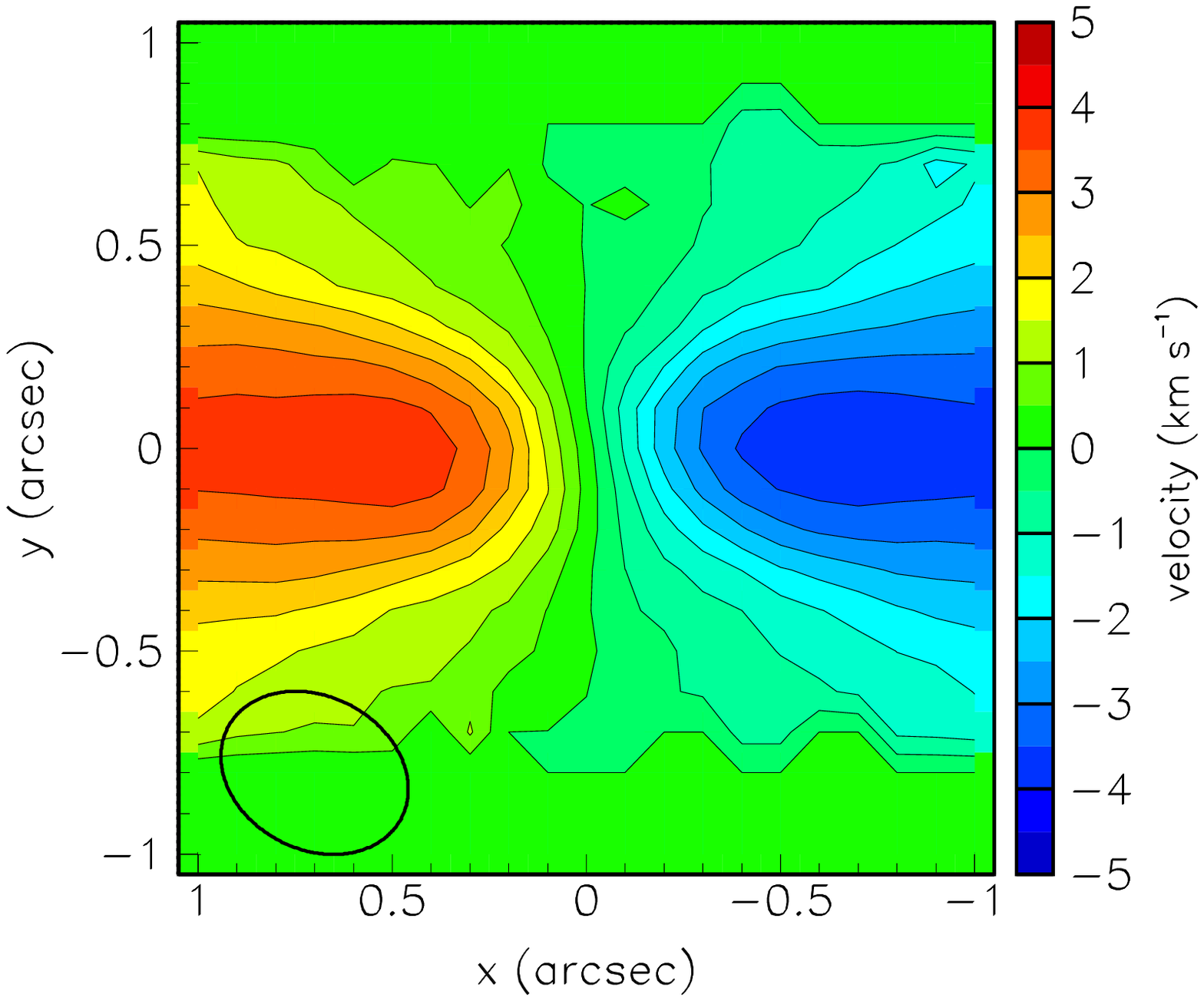}
\includegraphics[height=5.5cm,trim=0.cm 1.cm 1.cm 2.cm,clip]{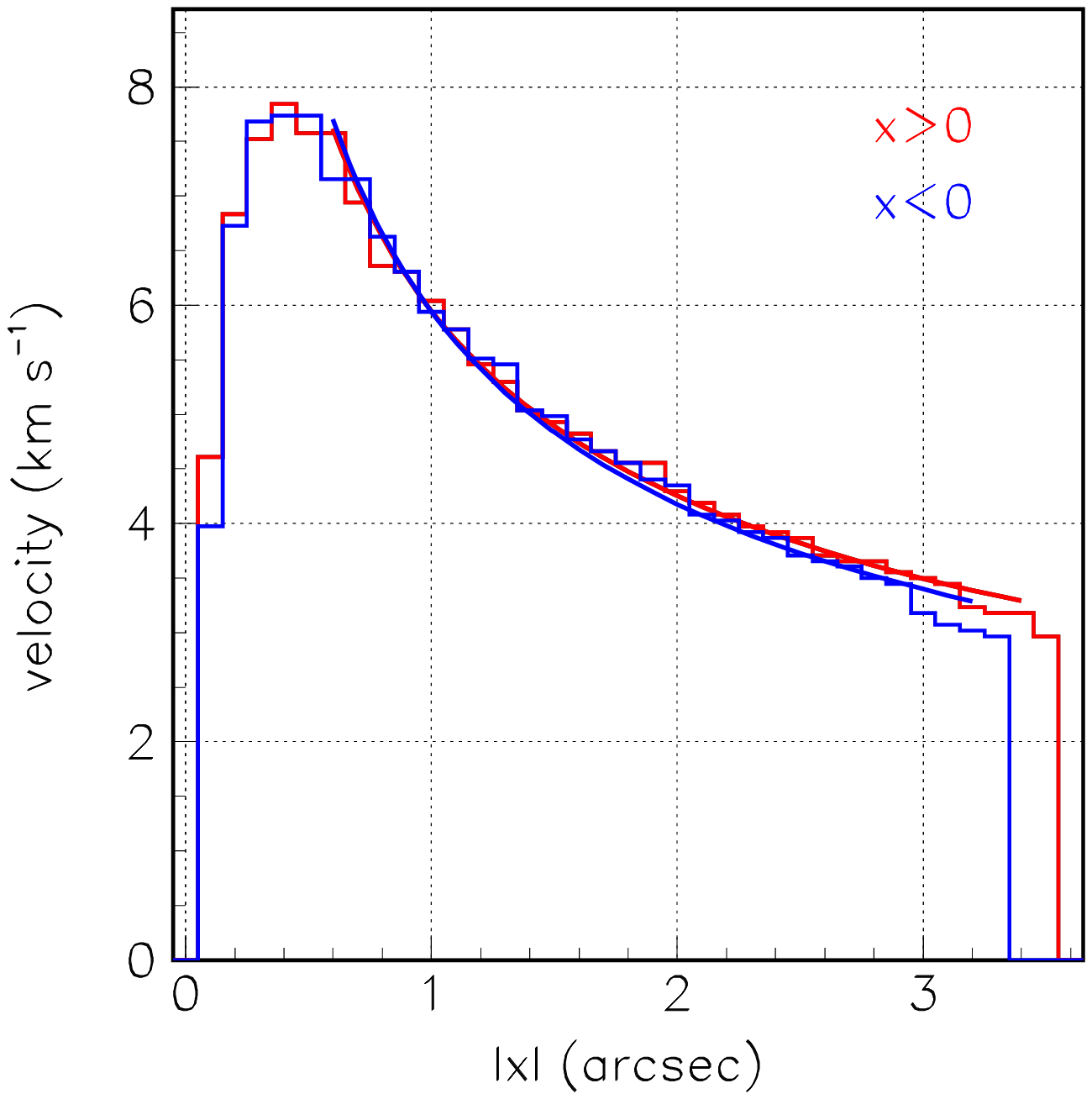}
\includegraphics[height=5.5cm,trim=0.cm 1.cm 1.cm 2.cm,clip]{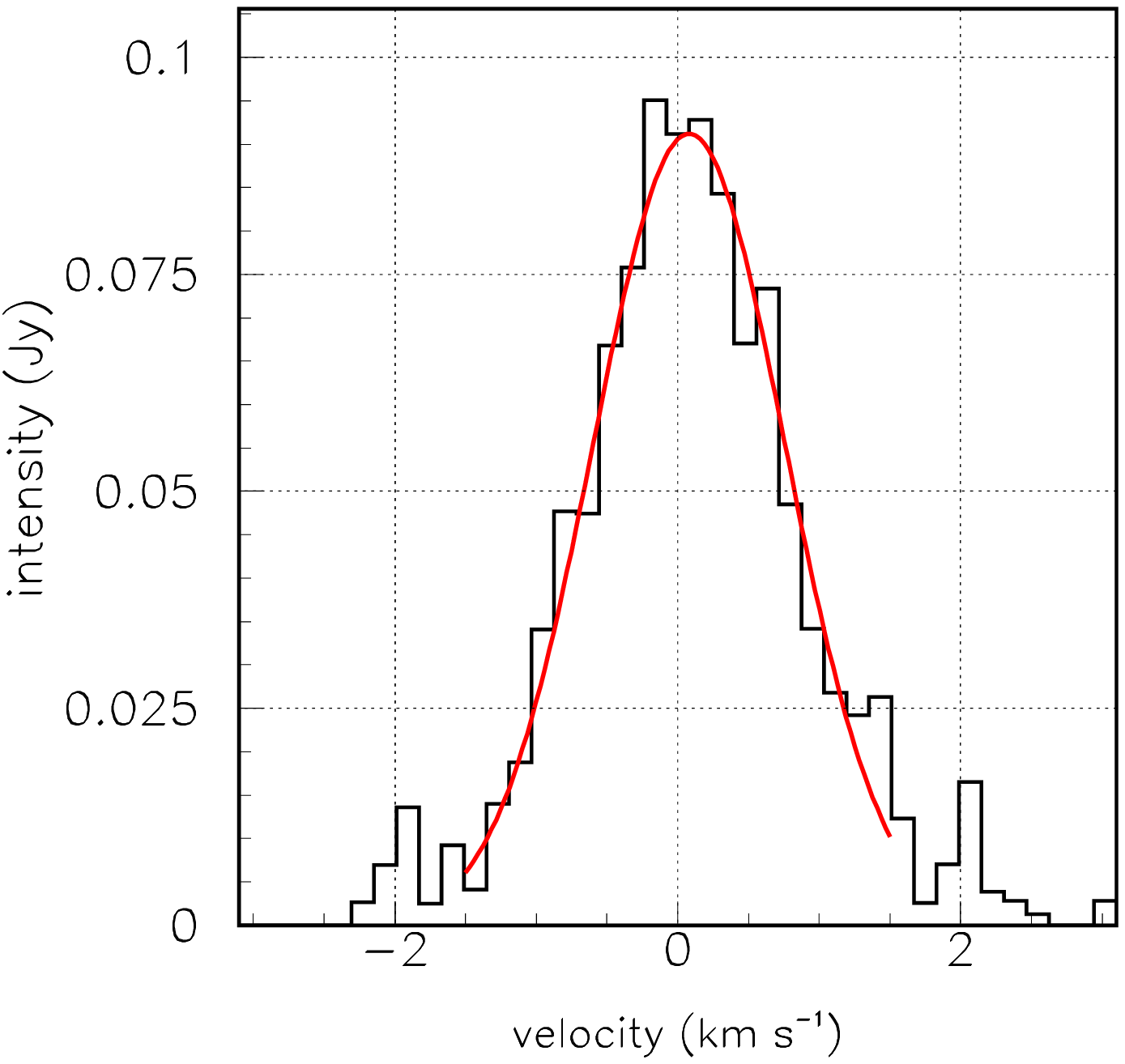}
\caption{Kinematics of CO(3-2) emission. Upper-left panel: measured distribution (blue) of the Doppler velocity for pixels having $\lvert x \rvert$<3.6 arcsec and $\lvert y \rvert$<1 arcsec. The red histogram is obtained by symmetry about the origin{, meaning that the flux density measured for a Doppler velocity $V_z$ is plotted at $-V_z$}. Upper-right panel: Map of the mean Doppler velocity in the {20$\times$20} central pixels. Lower-left panel: distribution of maximal absolute velocities (\kms) over $\lvert x \rvert$ for central ($y$=0) pixels. Red is for red-shifted ($x$>0) and blue for blue-shifted ($x$<0). The lines are power-law fits with respective power indices $-$0.47 and $-$0.51. Lower-right panel: {Distribution of intensity over Doppler velocity for the central ($x$=0) pixels.} The line is a Gaussian fit.}
\label{fig5}
\end{figure*}

Figure \ref{fig3} provides a more detailed picture, with $y$ profiles being displayed for different intervals of $x$. Remembering that both continuum and line data are observed with similar beam sizes (respectively 0.43 arcsec and 0.38 arcsec FWHM in $y$), we see that the emission of the dust disc extends significantly farther out in $y$ than the emission of the gas disc, with respective deconvolved FWHM values of \mbox{0.79 arcsec} and 0.53 arcsec. A priori, we do not know whether this difference is due to different inclination angles or to different extensions out of the disc mid-planes or both. Both distributions are very well centred on the origin, to better than \mbox{0.5 au}. The smooth regularity of the gas emission contrasts with that of $\beta$ Pictoris \citep{Dent2014}, with which 49 Ceti is often compared, and with the somewhat lumpy emission of the dust.

Figure \ref{fig4} shows that gas emission extends radially to only $\sim$3 arcsec while dust emission extends farther out, up to $\sim$4 arcsec. Correcting for the loss of acceptance associated with the {zero-spacing problem by using the Gaussian form of the acceptance introduced in Section 2 would increase the FWHM of the radial distributions of the line and continuum by 11$^{+6}_{-3}$\% and 18$^{+7}_{-5}$\% respectively, amounting to $\sim$0.3 and  $\sim$0.5 arcsec. The detailed morphologies of the central disc cavities are not immediately visible on the sky projections and the evaluation of their sizes requires de-projection of the images}; however, while the dust data display a broad and shallow central depression, as expected from an azimuthally symmetric disc, the gas data display a narrower and deeper central dip. Over the $x$ range, the mean values of the distributions shown in \mbox{Figure \ref{fig3}} fluctuate between $\pm$0.03 arcsec for the gas and \mbox{$\pm$0.05 arcsec} for the dust; the rms values fluctuate between 0.55 and \mbox{0.70 arcsec} for the gas and 0.80 and 1.05 arcsec for the dust.

\subsection{Gas kinematics}
The Doppler velocity (meaning the component $V_z$ of the velocity on the line of sight) spectrum, measured for CO(3-2) emission and integrated over the source, displays a double-horn structure, suggesting that the source is in rotation (Figure \ref{fig5}, upper-left panel). {The origin of the Doppler velocity scale is set at systemic velocity (2.78 \mbox{km s$^{-1}$}) and is discussed below.} Evidence for rotation is also clearly seen on the sky map of the mean Doppler velocity, of which the central part, {2$\times$2} arcsec$^2$, is displayed in the upper-right panel of Figure \ref{fig5}. The map is symmetric with respect to the $x$ axis and antisymmetric with respect to the $y$ axis and does not reveal any significant anomaly. However, typical line profiles of rotating discs display a central depression significantly shallower than observed here, with a peak to valley ratio of order 2 (see for example \citealt{Kospal2013} for the case of HD 21997); here, the peak to valley ratio is $\sim$5 and the valley, rather than being flat as for typical rotating discs, is deep and narrow over a velocity range of some 2 \kms. {Note that the relative depth of the valley is independent from inclination, all velocity bins being scaled by a common $\cos(i)$ factor.}

Figure \ref{fig5} (lower-left panel) displays the dependence on $x$ of the maximal absolute value of the Doppler velocity in the row of central pixels ($y$=0). Assuming circular orbits with velocities independent from azimuth, this is equal to the rotation velocity multiplied by the cosine of the inclination angle (close to unity). Very good fits to the distributions observed for $x$>0 and $x$<0 are obtained assuming a power law dependence of the velocity on radius, with respective values of the power index $n_{>0}$=$-$0.47 and $n_{<0}$=$-$0.51 and a rotation velocity of $\sim$6 \kms\ at a distance of 1 arcsec from the star. This provides excellent evidence for Keplerian ($n=-\frac{1}{2}$) rotation for radii in the [0.5 arcsec, 3.5 arcsec] interval (30 au to 210 au). At radii significantly smaller than $\sim$0.5 arcsec ($\sim$30 au), such a simple approach fails.

Figure \ref{fig5} (lower-right panel) displays the Doppler velocity distribution measured along the $y$ axis ($x$=0 column of pixels). A Gaussian fit excluding the small bumps near velocities of $\pm$2 \kms\ gives an rms value of \mbox{0.68 \kms} and a mean velocity of 0.08 \kms, meaning 2.86 \kms\ as the origin of the velocity scale is 2.78 \kms. It corresponds to 12.00 \kms\ in heliocentric coordinates when using an offset of 9.14 \kms\ to convert to the local standard of rest (LSR). We note that \citet{Hughes2008} use instead a value of 3.06 \kms\ and that \citet{Roberge2014} detect [C$_\textrm{I}$] at 2.45$\pm$0.40 \kms\ and [O$_\textrm{I}$] at \mbox{1.86$\pm$0.33 \kms} and \mbox{3.46$\pm$0.70 \kms}. A warm local cloud is present on the line of sight with a velocity of 1.86$\pm$1.3 \kms \citep{Redfield2008}.
In case of pure rotation on circular orbits, the distribution is expected to be centred at the origin and its dispersion should measure the spectral resolution, including broadening effects due to temperature and/or turbulence. The rms value of 0.68 \kms, meaning a FWHM value of 1.6 \kms, is equal to the Doppler broadening parameters measured at FUV frequencies by \citet{Roberge2014} for [C$_\textrm{I}$] and [O$_\textrm{I}$] between 1.3 and 1.8 \kms. Deviations from such behaviour would signal radial velocities such as outflows or in-fall. The small bumps near velocities of $\pm$2 \kms\ are associated with $y\sim$0.

\begin{figure*}
\centering
\includegraphics[height=6.cm,trim=0.cm 0.cm 0.5cm 2.cm,clip]{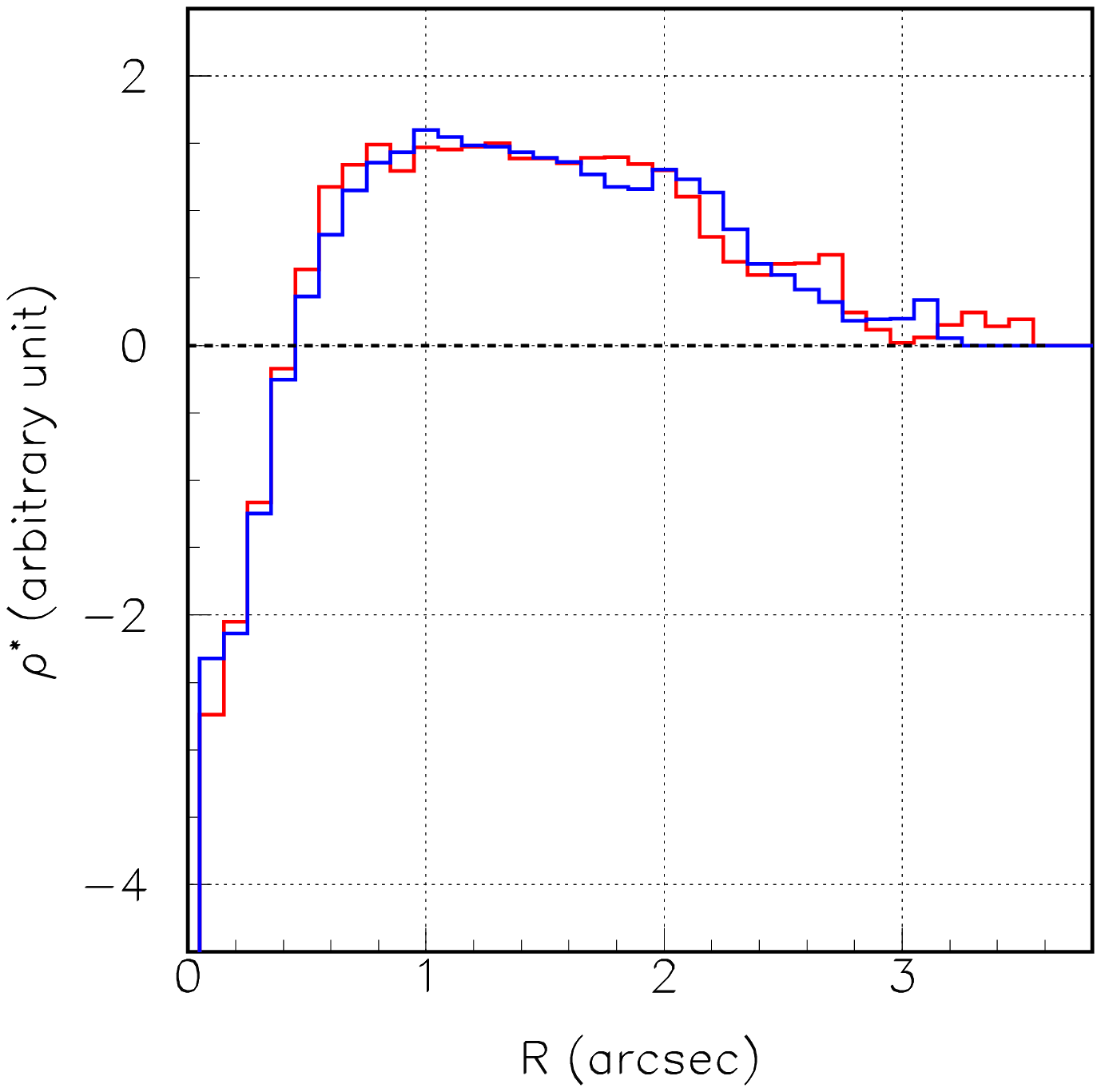}
\includegraphics[height=6.cm,trim=0.cm 0.cm 1.0cm 2.cm,clip]{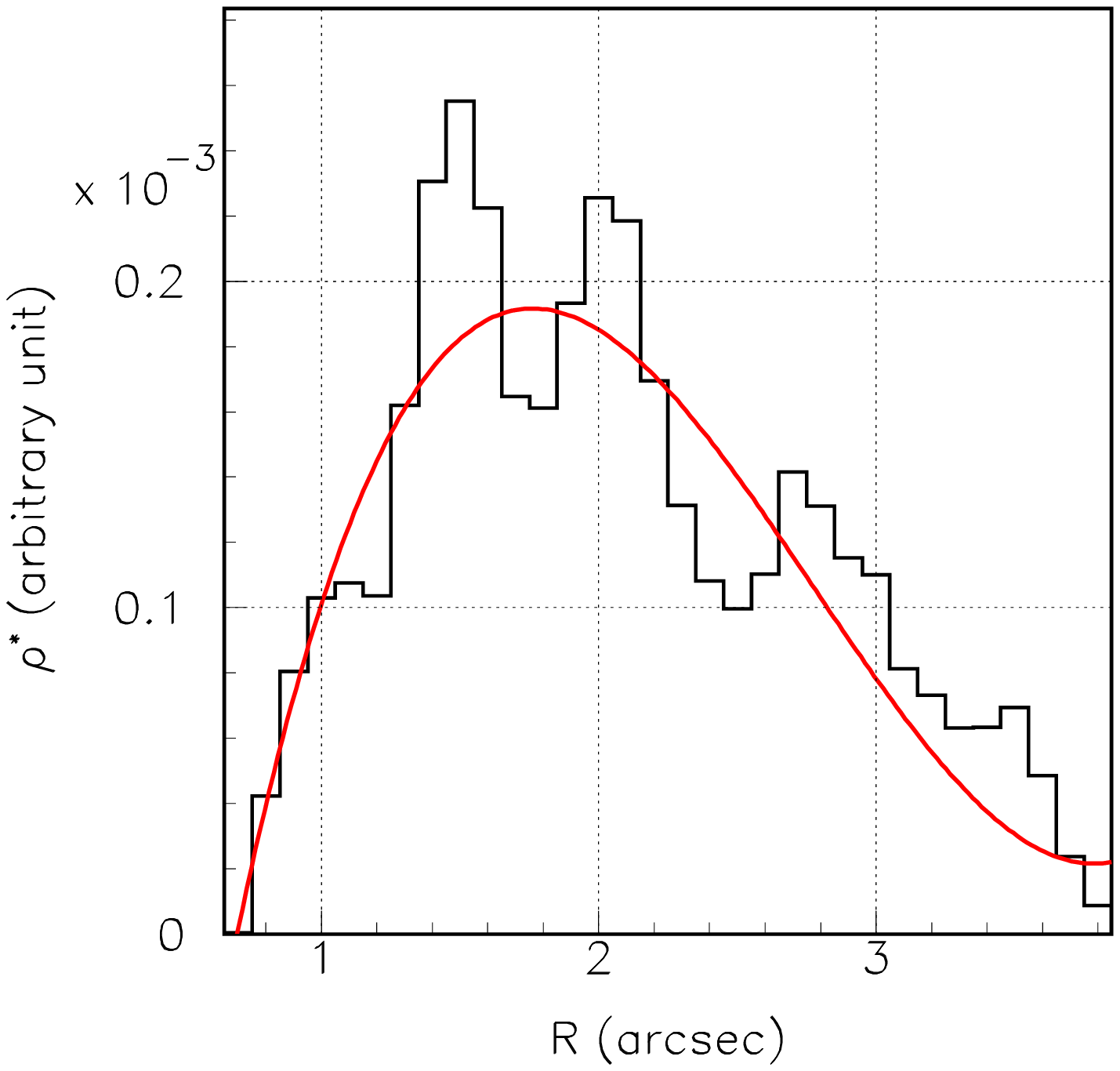}
\caption{Radial dependence of the effective emissivity calculated under the hypotheses of optical thinness and invariance by rotation about the disc axis. Left: CO(3-2) emission; the red histogram is obtained from the $x$>0 data, the blue histogram from the $x$<0 data. Right: 350 GHz continuum (summing $x$<0 and $x$>0); the line is the third degree polynomial fit used in the model (note the offset zero of the $R$ axis). In both panels we use arbitrary units to measure the effective emissivity as only its shape matters here.}
\label{fig6}
\end{figure*}

\section{Properties of the CO(3-2) emission}
\subsection{Integrated intensity of the CO(3-2) emission}
We evaluate the integrated intensity of the \mbox{CO(3-2)} emission by summing over all pixels exceeding 3 $\sigma$'s, accounting for the Gaussian beam profile and correcting for lower contributions by extrapolating under the noise using the distribution displayed in Figure \ref{fig1}, giving 7.1$\pm$0.4 Jy \kms. The correction for the loss of short spacing cannot be reliably evaluated but is estimated, as described in Section 2, at the level of +1.6 Jy \kms. In addition, a normalization uncertainty of $\sim$10\% associated with the calibration of the absolute flux must be taken in account. The result is in good agreement with the single dish value of 9.5$\pm$1.9 Jy \kms\ obtained by \citet{Dent2005} using the James Clerk Maxwell Telescope (JCMT). \citet{Hughes2008} using SMA observations have detected CO(2-1) emission at the level of 2.0$\pm$0.3 Jy \kms.

Under the hypothesis of optical thinness and local thermal equilibrium, \citet{Hughes2008}, using the proportionality relation between integrated intensity and gas mass, find that the CO(2-1) SMA measurement corresponds to a CO mass of 2.2$\times$10$^{-4}$ Earth masses. Using the same relation, with coefficients adapted to CO(3-2) emission and assuming a temperature of 50$\pm$10 K as suggested by \citet{Hughes2008}, our measured intensity of  7.1$\pm$0.4 Jy \kms\, translates into a CO mass of 2.2$\times$10$^{-4}$ Earth masses that would increase to $\sim$2.7$\times$10$^{-4}$ Earth masses using our crude estimate of the short-spacing correction.
\mbox{\citet{Hughes2008}}, using a model of thermal and chemical processes, find that the surface temperature of the disc increases from $\sim$40 K to $\sim$60 K when the distance to the star decreases from 200 au to 50 au, and increases by a factor 3 to 5 when moving across the disc from surface to mid-plane, implying optical thinness with axial CO column densities not exceeding a few 10$^{15}$ molecules per square centimetre. Note that under the hypothesis of CO being outgassed by colliding comet-like bodies, the CO mass cannot be used to obtain a total gas mass, the CO/H$_2$ ratio being unconstrained and probably orders of magnitude larger than the canonical value of 10$^{-4}$.

\begin{figure*}
\centering
\includegraphics[height=6.cm,trim=-.5cm 1.cm 0.cm 2.cm,clip]{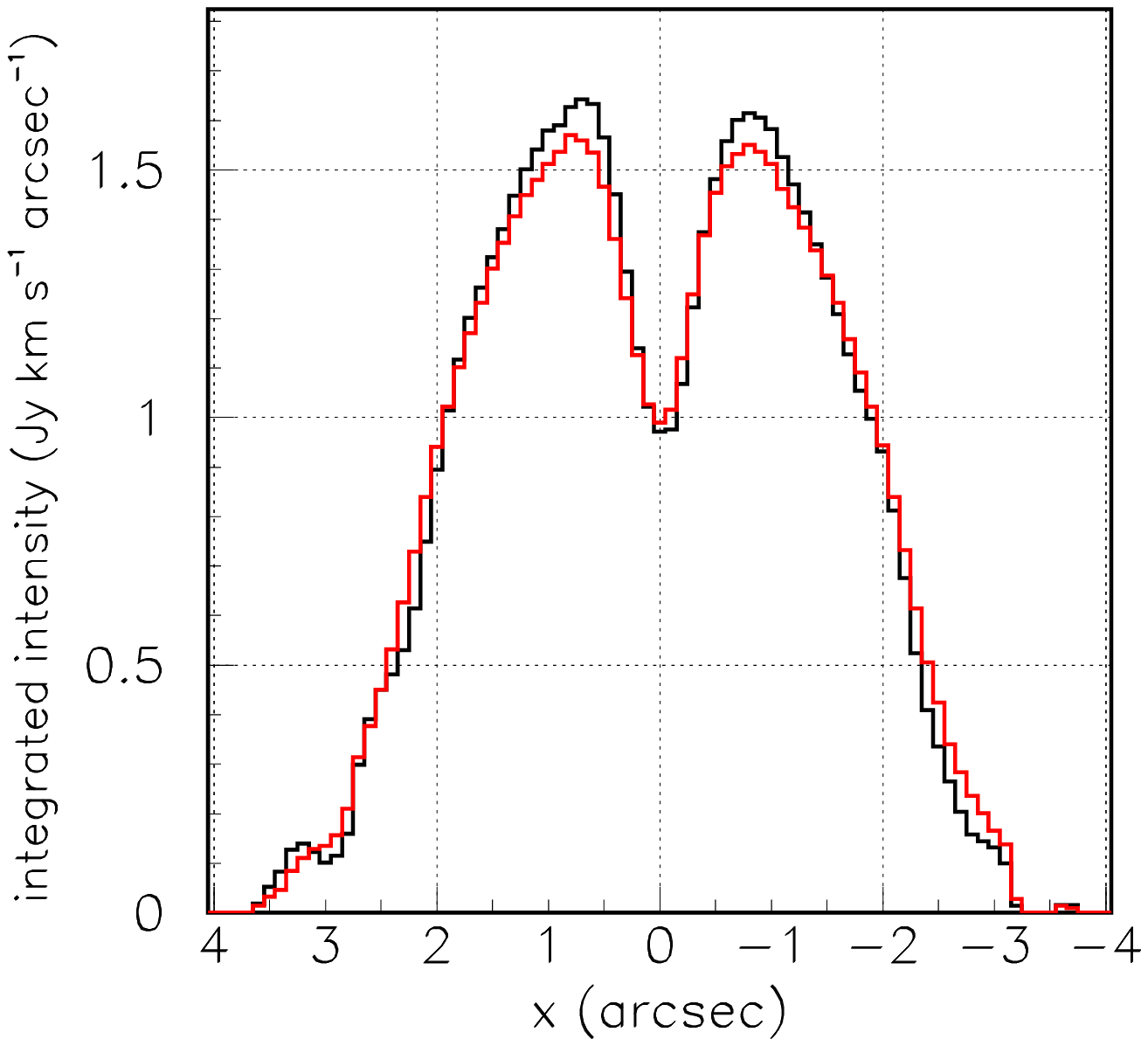}
\includegraphics[height=6.cm,trim=0.cm 1.cm 0.cm 2.cm,clip]{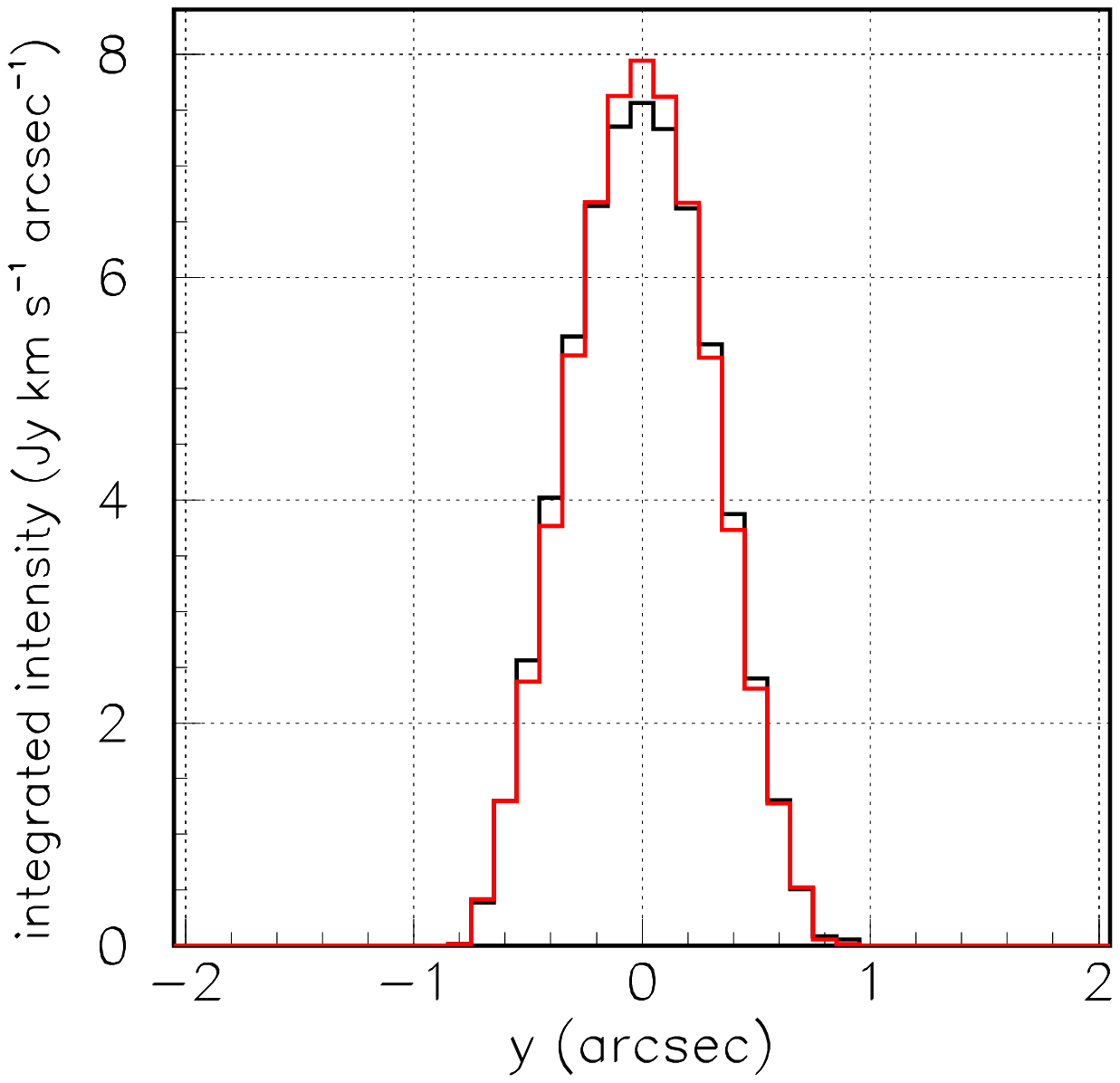}
\includegraphics[height=6.cm,trim=0.cm 1.cm 0.cm 2.cm,clip]{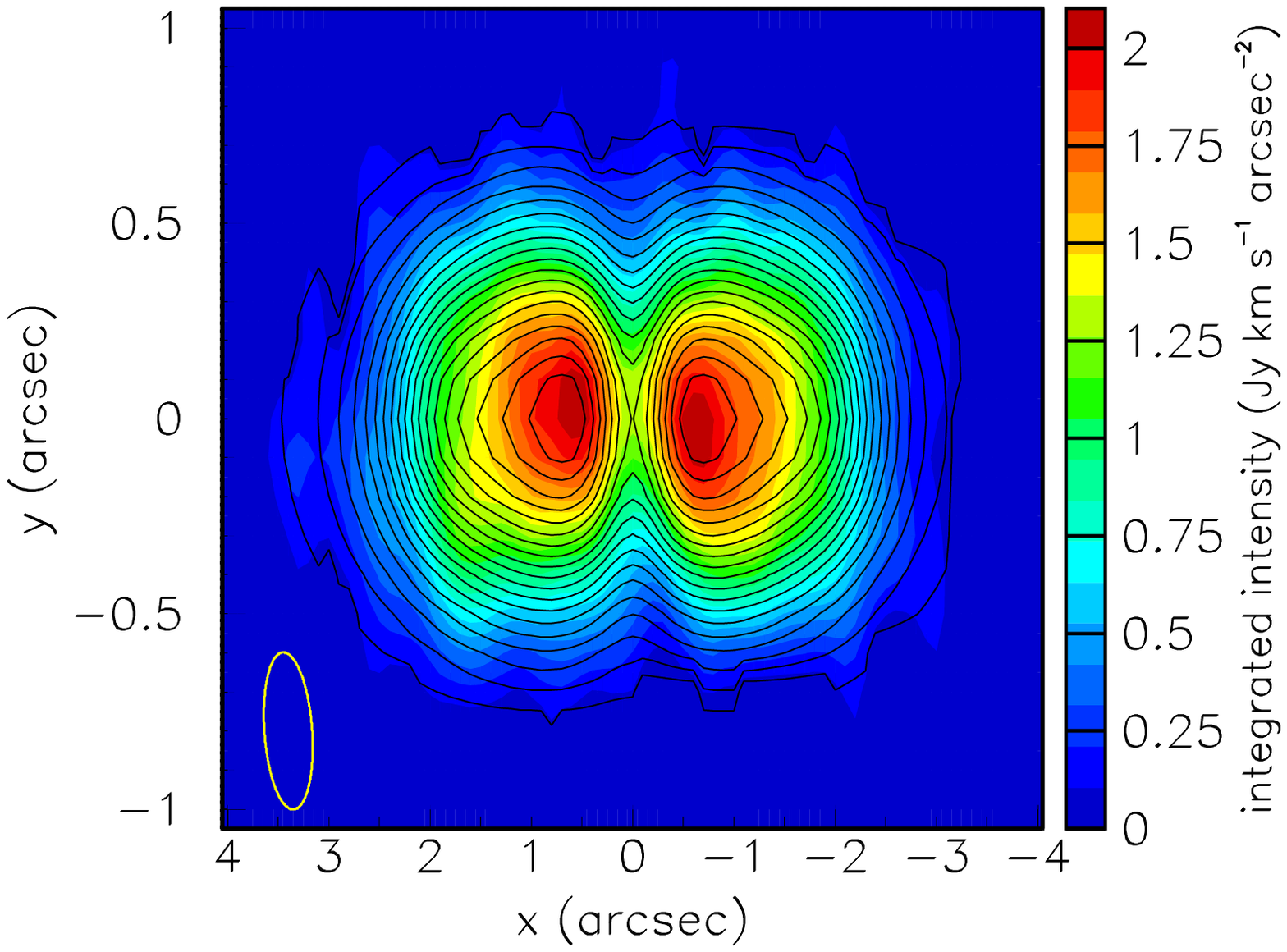}
\includegraphics[height=6.cm,trim=-0.5cm 0.3cm 0.cm 1.7cm,clip]{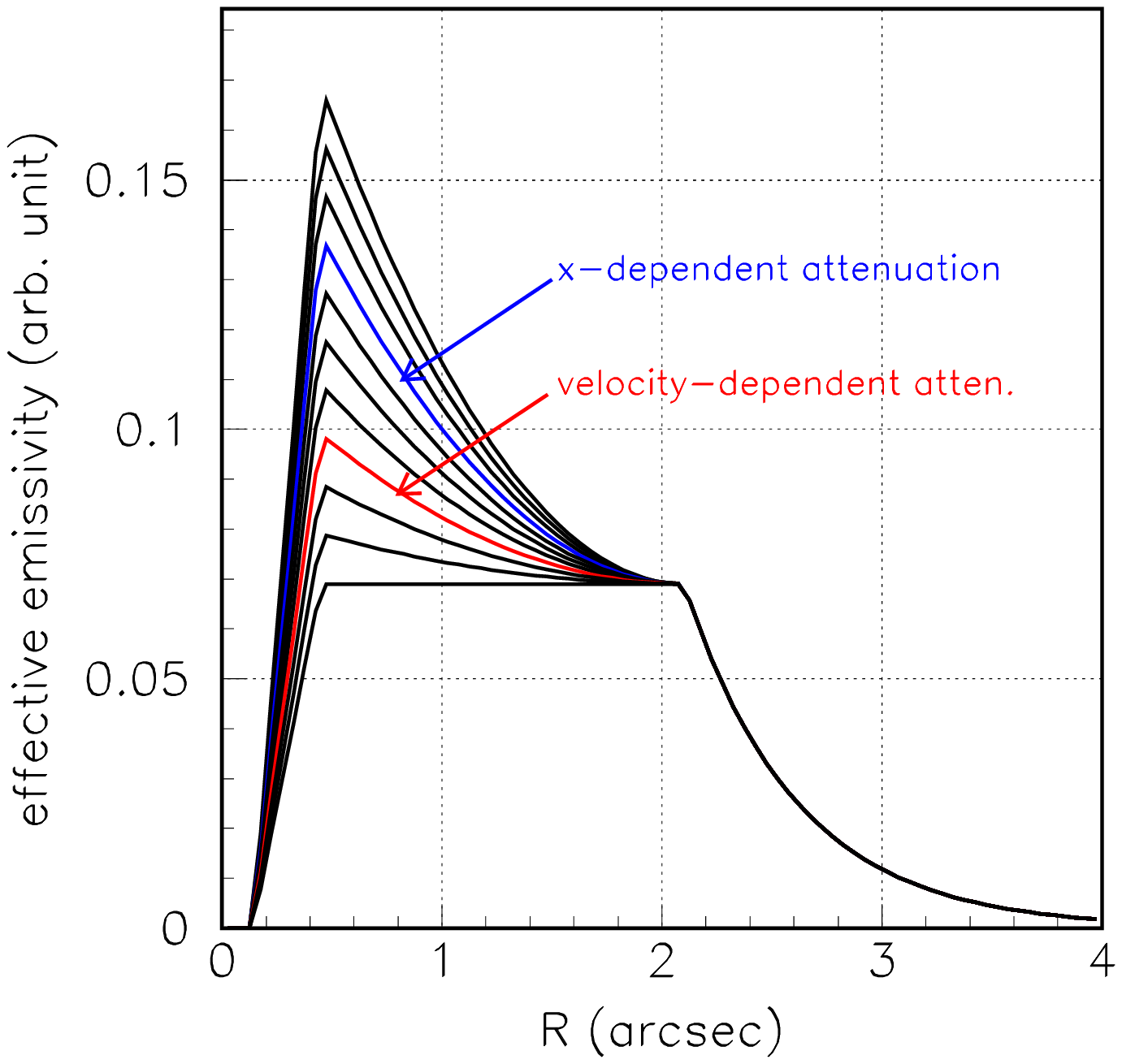}
\caption{Morphology of CO(3-2) emission. Upper panels: $x$-distribution (left) and $y$-distribution (right) of $F(x,y)$ integrated over $y$ and $x$ respectively. The red histograms are the results of the best fit. Lower left panel: map of the velocity integrated intensity as measured (colour) and as obtained from the best fit (contours). {Note the different $x$ and $y$ scales.} Lower right panel: family of curves describing the radial dependence of the effective emissivity (before imposing attenuation); the index $k$ varies from 0 to 1 from lower to higher emissivity; the best fit curves are shown in red (velocity-dependent attenuation) and in blue ($x$-dependent attenuation).}
\label{fig7}
\end{figure*}

\subsection{Evidence for central attenuation of the CO(3-2) emission}
{When attempting to model the line emission as described in the next section, we fail to adjust the parameters of the model to reproduce the narrow central dip observed in Figure \ref{fig4}. The present section aims at giving in simple terms a clear picture of the reason behind it, leaving for the next section a more rigorous treatment.} Indeed no optically thin model assuming rotation invariance about the disc axis can do so; to show it, we first remark that under such assumptions the inclination of the disc, its thickness, whether it is flared or not and beam convolution are to first order irrelevant to the argument. The point is simply that the emission observed at large values of $\lvert x \rvert$, namely at large disc radii, contributes more than what is observed along lines of sight pointing to $x=0$.

We define \citep{Diep2016} the effective emissivity $\rho(x,y,z)$ as
\begin{equation}
\centering
\rho(x,y,z)=f(x,y,V_z)\frac{dV_z}{dz}
\end{equation}
where $f(x,y,V_z)$ is the flux density {and $V_z$ the Doppler velocity defined earlier}. It combines the contributions of density, temperature and optical thickness in a single quantity at each point in space.
It implies that the intensity measured in a pixel $(x,y)$ reads
\begin{equation}
\centering
F(x,y)=\int f(x,y,V_z)\,dV_z=\int\,\rho(x,y,z)\,dz
\end{equation}
Neglecting disc inclination and disc thickness, and assuming that the effective emissivity is invariant by rotation about the disc axis (in practice it means that temperature and density are invariant by rotation about the disc axis and that the disc is optically thin)  the intensity measured in an interval $dx$ reads, after integration over $y$
\begin{equation}
\centering
F^*(x)\,dx=\int \rho^*(R)\,dz
\end{equation}
with $F^*$ and $\rho^*$ being the integrals over $y$ of $F$ and $\rho$ and \mbox{$R=\sqrt{x^2+z^2}$} measuring the radius in the disc plane. This integral equation is easily solved by steps from large $\lvert x \rvert $ values inward; the result is displayed in Figure \ref{fig6} (left) and shows that $\rho^*(R)$ needs to become negative for \mbox{$R$<$\sim$0.5 arcsec} in order to compensate for too strong emission at large distances from the star. This important result demonstrates that UV dissociation of the CO molecules in the vicinity of the star is not sufficient to account for the deep and narrow depression of the measured CO(3-2) emission at small values of $\lvert x \rvert$. 
{The assumption of invariance of the effective emissivity by rotation about the disc axis needs to be given up}, allowing for significant attenuation of the detected emission near $x$=0 (covering at least \mbox{$\sim$1 arcsec} FWHM, meaning $\sim$60 au, in $x$). {This may mean that either optical thinness or azimuthal symmetry of the temperature and/or density distributions are violated.}

Guided by this result, we describe the $x$-dependence of the measured intensity using a Gaussian form \mbox{$A(x)=1-\lambda_x\,\textrm{exp}(-\frac{1}{2}x^2/\Delta_x^2)$} to model the observed central attenuation. {Such a factor provides the required azimuthal symmetry breaking in the simplest possible way.} For any given value of $\Delta_x$, we divide the {observed intensity} by $A(x)$ and adjust $\lambda_x$ to have the solution of integral equation (3) give an effective emissivity that cancels exactly at $R=0$, as it should. We obtain this way, for each value of $\Delta_x$, the corresponding value of $\lambda_x$ and the associated radial dependence of the effective emissivity. The latter are displayed in Figure \ref{fig7} (lower right) as a one-parameter family of functions of an index $k$ simply related to $\Delta_x$.

\subsection{A simple model of CO(3-2) emission}
{In the present section we aim at giving as simple as possible a model of the morphology and kinematics of the debris disc of 49 Ceti. The very high quality of the data makes it possible to do so more reliably than was previously possible. Our approach is therefore to use as little parameters as possible, namely the minimal set required to describe a disc: position angle, inclination, flaring angle and radial distribution. Of these the position angle, which is not correlated to the other parameters, has been evaluated earlier and does not need to be further adjusted. The radial distribution is in principle obtained from the solution of integral equation (3); however, the presence of an unexpected central attenuation, together with the fact that integral equation (3) ignores the effect of beam convolution, suggests giving some additional freedom to the radial distribution; we do so by using a one-parameter family of functions as described at the end of the preceding section and illustrated in Figure \ref{fig7} (lower-right panel). Moreover, the presence of a central attenuation implies an additional factor in the model, for which we propose two different forms, each requiring the introduction of two additional parameters. The real justification of the adopted approach is the excellent quality of the fits obtained to the observed distributions. Such a pragmatic approach favours working in the image plane rather than in the uv plane and dealing with the effective emissivity rather than independently with temperature, density and chemical composition. The model integrates the modelled flux density over the line of sight and accounts for beam convolution. Precisely, as argued and described below, three sets of data are jointly used in the evaluation of $\chi^2$:  the sky maps of the velocity integrated intensity and of the mean velocity, and the global velocity spectrum.}

The regularity of the map of CO(3-2) emission and its nearly perfect symmetry about the $y$ axis make it unlikely that rotation invariance about the disc axis be significantly violated. Therefore, in constructing a model of the morphology and kinematics of CO(3-2) emission, we assume rotation invariance about the disc axis and symmetry about the disc mid-plane. We also assume that the disc mid-plane is flat (no warping) and that the disc is flared. {However, as explained in the preceding section, these assumptions are somehow violated. In the model, we account for this violation by introducing an ad-hoc central attenuation that violates rotational invariance, leaving for later a possible physical interpretation.} An important constraint on the model is the need for the CO column density along the line of sight pointing to the star not to exceed $\sim$5$\times$10$^{12}$ cm$^{-2}$ in order to account for the HST observations of \citet{Roberge2014} but, in any case, the thickness of the disc cannot be large: the deconvolved value of the FWHM of the $y$ distributions displayed in Figure \ref{fig3}, 0.53 arcsec, is easily accounted for by an inclination of the order of 10$^\circ$, leaving little room for the disc thickness to deviate from 0. We find it reasonable, under such conditions to write the dependence of the effective emissivity on the distance $h$ to the disc mid-plane as   
\begin{equation}
  R^{-1}\textrm{exp}(-\frac{1}{2}h^2/\Delta_h^2) \, \, \, \, \textrm{with}\, \, \Delta_h=\eta_hR
\end{equation}
where $R$, the radius measured in the disc plane, reads now $R=y/\textrm{sin}(i)-h/\textrm{tan}(i)$, and $\eta_h$ is a parameter to be adjusted. The choice of a Gaussian form follows \citet{Hughes2008} and is not critical, the smearing resulting from beam convolution being relatively important.

\begin{figure*}
\centering
\includegraphics[height=4.2cm,trim=2.cm 1.cm 0.cm 0.cm,clip]{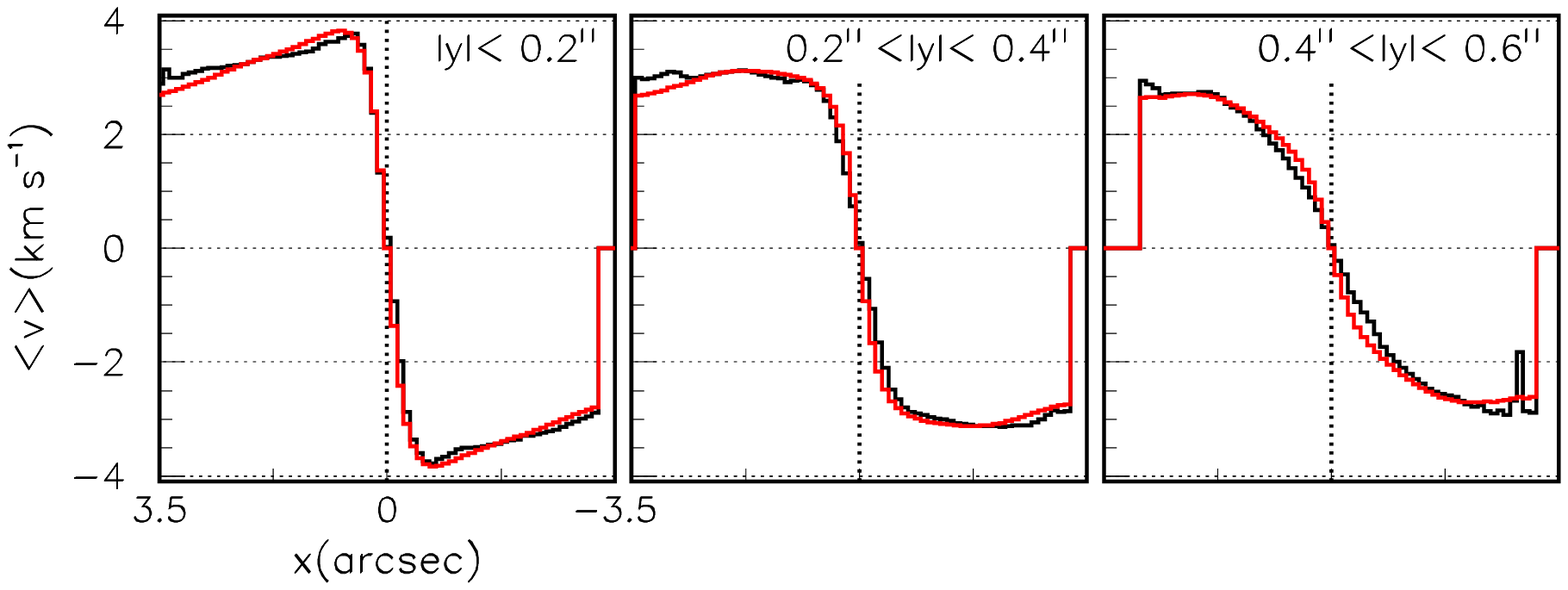}
\includegraphics[height=6.cm,trim=0.cm 1.cm 0.cm 2.cm,clip]{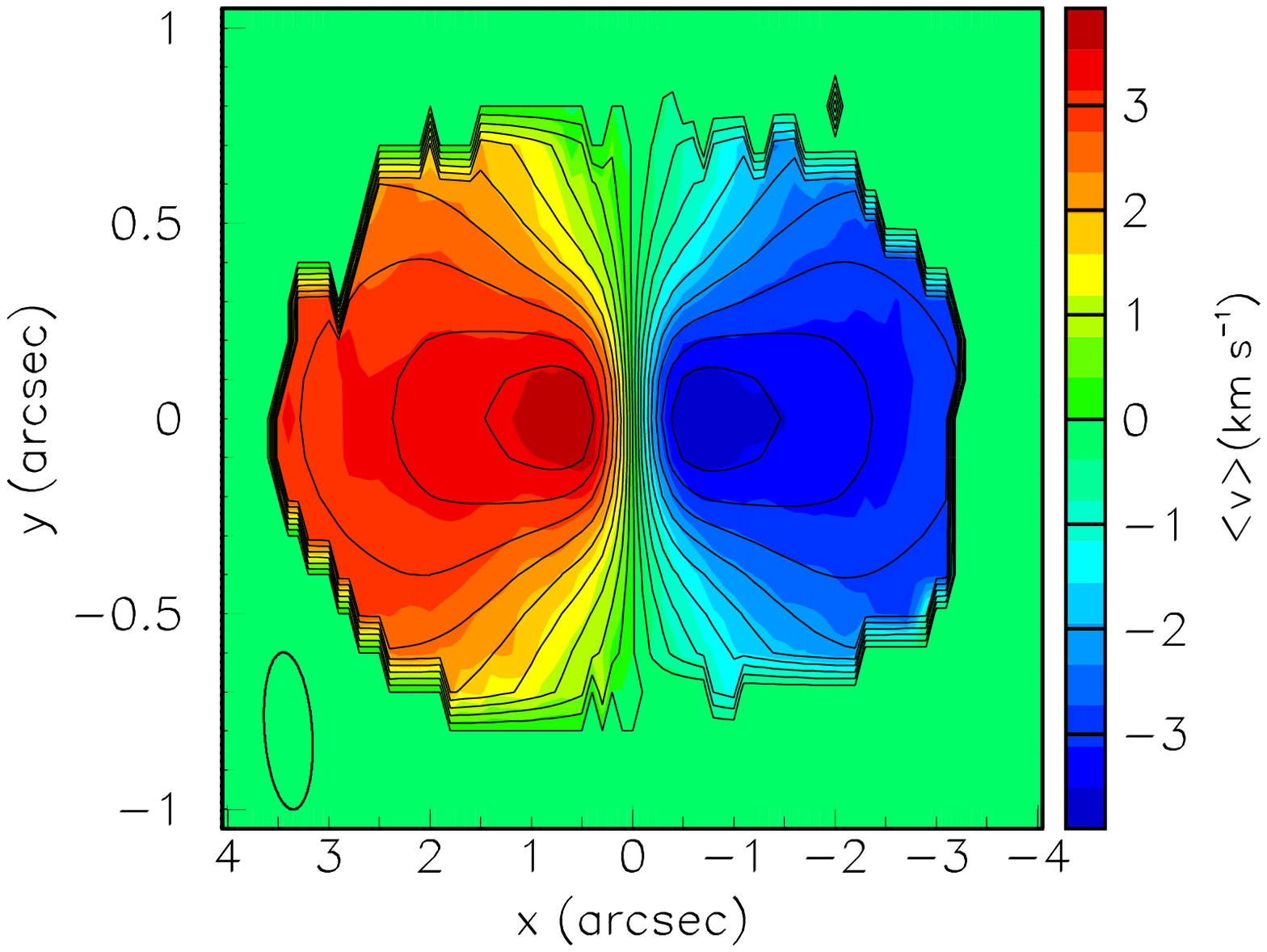}
\includegraphics[height=6.cm,trim=0.cm 1.cm 0.cm 2.cm,clip]{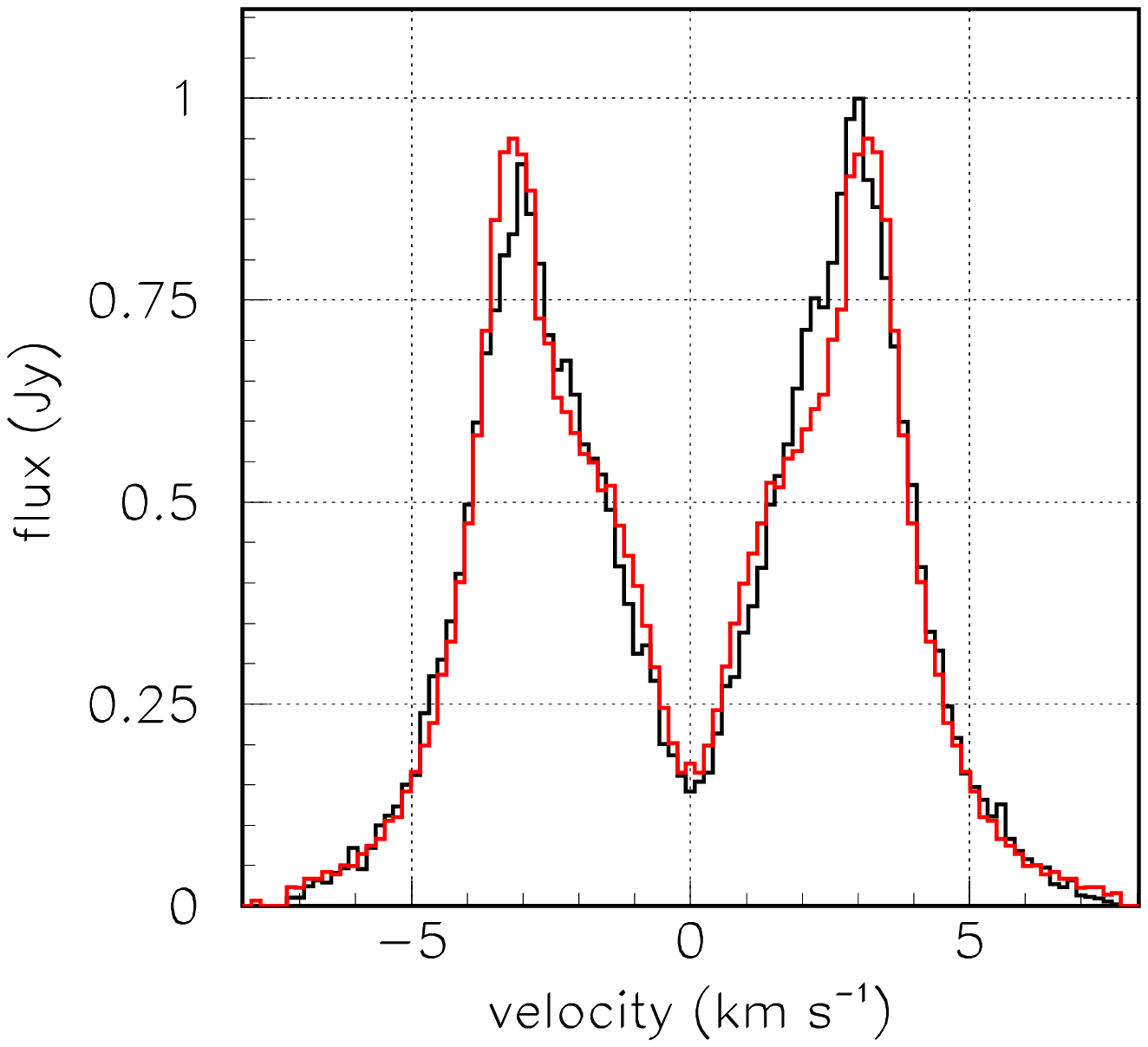}
\caption{Kinematics of CO(3-2) emission. Upper panels: $x$-distributions of the mean Doppler velocity for $\lvert y \rvert$<0.2 arcsec (left), \mbox{0.2 arcsec<$\lvert y \rvert$<0.4 arcsec} (middle) and 0.4 arcsec<$\lvert y \rvert$<0.6 arcsec (right). The black histograms display the data and the red histograms display the result of the best fit. Lower left panel: map of the mean Doppler velocity as measured (colour) and as obtained from the best fit (contours); note the different $x$ and $y$ scales. Lower right panel: mean Doppler velocity distribution for data (black) and best fit (red).}
\label{fig8}
\end{figure*}

\begin{table*}
\centering\setcellgapes{2pt}\makegapedcells \renewcommand\theadfont{\normalsize\bfseries}%
\caption{Best fit parameters of a simple disc model of CO(3-2) emission allowing for $V_z$-dependent ({left-hand} values) or $x$-dependent ({right-hand} values) attenuation. The quoted uncertainties are for a 10\% increase of $\chi^2$, leaving all other parameters free to vary.}
\label{Table 1}        
\begin{tabular}{|c|c|c|c|}
\hline 
Description & Name & \multicolumn{2}{c|}{Value} \\
\hline
\multicolumn{4}{|c|}{Intensity morphology} \\
\hline
Inclination & $i$ (degrees) & \multicolumn{2}{c|}{$11.1\pm0.9$} \\
\hline
Flaring angle & $\eta_h$ (rad) & \multicolumn{2}{c|}{$0.041^{+0.028}_{-0.019}$} \\
\hline
Radial dependence & $k$ & $      0.29^{+0.31}_{-0.22}$ & $0.70^{+0.37}_{-0.22}$     \\
\hline
\multicolumn{4}{|c|}{Rotation velocity} \\
\hline
Velocity at $r$=1 arcsec & $V_0$ (\kms) & \multicolumn{2}{c|}{$5.1\pm0.1$} \\
\hline
Power index & $n$ & \multicolumn{2}{c|}{$-0.46^{+0.05}_{-0.02}$} \\
\hline
\multicolumn{4}{|c|}{Attenuation at small $\lvert V_z \rvert$ or small $\lvert x \rvert$ values} \\
\hline
Amplitude & $\lambda$ & 0.66$\pm$0.15 & 0.60$\pm$0.12 \\
\hline
\multirow{1}{*}{FWHM in $V_z$ or $x$} & \makecell{2.355 $\Delta_V$ (\kms) \\or  2.355 $\Delta_x$ (au)}  & \multirow{1}{*}{1.5$\pm$0.5} & \multirow{1}{*}{49$\pm$17} \\
\hline
\multicolumn{2}{|c|}{$\chi^2$ per degree of freedom}& 0.76 & 0.80 \\
\hline
\end{tabular} 

\end{table*}

The radial dependence of the effective emissivity is taken from the family of radial functions obtained at the end of Section 4.2 and displayed in Figure \ref{fig7} (lower right panel). We recall that correcting for the loss of acceptance at large values of $\lvert x \rvert$ increases the radial extension of these functions by $\sim$0.2 arcsec.

In addition to the sky map of the velocity integrated intensity, we also fit the sky map of the mean Doppler velocity and the Doppler velocity spectrum itself. As we want to maintain good spectral resolution and large enough signal-to-noise ratio, we find this more efficient than fitting each individual data-cube element separately, at the price, however, of losing information on the detailed dependence of the velocity profiles over the sky. We model the rotation velocity as $V_0r^n$, where $V_0$ and $n$ are two new parameters to be adjusted and where $r$ is the distance to the star, $r=\sqrt{x^2+y^2+z^2}=\sqrt{R^2+h^2}$. The morphology and kinematics are therefore described by five parameters: the inclination $i$ of the disc with respect to the line of sight, the index $k$ defining the radial dependence of the effective emissivity, the flare parameter $\eta_h$ describing the $r$-dependence of the disc thickness and the velocity parameters $V_0$ and $n$.

We are then left with the problem of modelling the observed attenuation of the CO(3-2) emission in some ad-hoc way.

The evidence obtained from mid-infrared observations \citep{Wahhaj2007} for a warm and fine-grained dust component confined to the vicinity of the star suggests associating the attenuation with this region. Indeed, commenting on their recent HST observation of the star, \citet{Roberge2014} mention the possible presence of a central disc orbiting the star, thicker than the molecular gas disc and deprived of molecular gas but rich in atomic species. This suggests a parameterization of the attenuation of the form
\begin{equation}
  A(x,y)=1-\lambda\,exp(-[x^2/\Delta_x^2+y^2/\Delta_y^2])
\end{equation}
with $\lambda$ measuring the amplitude of the attenuation and $\Delta_x$ and $\Delta_y$ measuring the extensions of the attenuation region in $x$ and $y$ respectively. However, we find that good fits require $\Delta_y$ to exceed $\Delta_x$, which is very surprising: if the central mid-infrared emission is from a disc having the same inclination as the gas disc, one would expect $\Delta_y$ to be significantly smaller than $\Delta_x$; the best fit to the CO(3-2) attenuation requires instead that it be at least as large as $\Delta_x$. More precisely, \citet{Wahhaj2007} quote an inclination of the warm disc of 55$^\circ\pm20^\circ$. If we insist to associate the observed \mbox{CO(3-2)} attenuation with this disc, we expect $\Delta_y/\Delta_x$ $\sim$0.5$\pm$0.3, allowing indeed for large $\Delta_y$ values, however smaller than $\Delta_x$. The radial extension of the molecular disc along the y axis is only $\sim$3$\times$sin(11$^\circ$)=0.6 arcsec \mbox{($\sim$40 au)}. What the result really implies is therefore simply that the data give no evidence for any $y$ dependence of the attenuation. Accordingly, we simply model the attenuation as
\begin{equation}
  A(x)=1-\lambda_x\,exp(-\frac{1}{2}x^2/\Delta_x^2), 
\end{equation}
implying two additional parameters to be adjusted, $\lambda_x$ and $\Delta_x$. However, in practice, it is very difficult to imagine a morphology of the attenuation region that could be independent of $y$. The point is that the attenuation needs to affect the whole $x\sim 0$ emission, including at large disc radii. Moreover it must do so for $y>0$ as well as for $y<0$, in spite of the fact that one of these regions corresponds to the half of the CO disc in the foreground of the star and the other to the other half of the CO disc, in the background of the star.

This difficulty suggests modelling the attenuation as a function of frequency rather than as a function of $x$. Indeed, the present situation is reminiscent of young proto-planetary discs observed behind a cloud of in-falling gas (e.g. \citealt{Tuan-Anh2016}). The difference is that the present attenuation displays no Doppler shift with respect to the disc rest frame, meaning negligible in-fall velocity, and, of course, that the in-falling gas that formed the 40 Myr old 49 Ceti should have disappeared since long. An important asset of the new parameterization is that it naturally reproduces the shape of the region of the sky affected by attenuation, its extension to large values of $\lvert y \rvert$ being no longer a problem. We can therefore replace the form (6) used to model the attenuation by a form
\begin{equation}
  A(V_z)=1-\lambda_V\,\textrm{exp}(-\frac{1}{2}V_z^2/\Delta_V^2)  
  \end{equation}
with parameters $\lambda_V$ and $\Delta_V$ replacing $\lambda_x$ and $\Delta_x$.

The model predictions are convolved with the beam and compared with the observed values, the parameters being adjusted to minimize $\chi^2$. We use an uncertainty of \mbox{0.22 \kms} on the mean Doppler velocities and of \mbox{14.6 mJy beam$^{-1}$ \kms} (1-$\sigma$) added in quadrature to a 10\% relative uncertainty on the velocity integrated intensity. There is some arbitrariness in these definitions; we scaled the uncertainties in such a way to have each of the intensity map, mean velocity map and velocity spectrum contribute approximately equal amounts to the total $\chi^2$.  We checked that these choices have negligible influence on the results. In practice we minimize $\chi^2 = \chi^2_1+\chi^2_2+\chi^2_3$ with each term defined as the square of the difference between data and model divided by the square of the relevant uncertainty as defined above. Here $\chi^2_1$ is for the sky (clean) map of the velocity integrated intensity, summed over its value in each of the 80$\times$20 pixels; $\chi^2_2$ is similarly for the sky map of the mean Doppler velocity; $\chi^2_3$ is for the Doppler velocity spectrum, summed over its value in each of 32 velocity bins. The minimization of $\chi^2$ is straightforward and uses FORTRAN code MINUIT \citep{James1975}. 

Given the simplicity of the model, very good fits are obtained in either case, $V_z$-dependent or $x$-dependent attenuation. They are illustrated in Figures \ref{fig7} and \ref{fig8} and the best fit parameters are listed in Table \ref{Table 1}. The best fit radial dependence of the effective emissivity (Figure \ref{fig7}, lower-right) cancels below radii of 10 to 15 au, then rises sharply before decreasing with radius from $\sim$20 to 25 au onward.  The inclination of the disc, $\sim$11$^\circ$, is better constrained by the data than the opening angle of the flaring, 5$^\circ$ to 6$^\circ$ FWHM. Their values comfortably satisfy the need to leave a CO-free region on the line of sight pointing to the star, as required by the HST data of \citet{Roberge2014}.

In both cases, attenuation is important, of the order of 2/3 at maximum. Its extension in $x$, $\sim$50 au FWHM, is smaller than that obtained by \citet{Wahhaj2007}, \mbox{120$\pm$40 au}, for the central warm disc. Its extension in $V_z$ is 1.5 \kms\ FWHM, significantly larger than the spectral resolution. It is remarkable that both attenuation models produce nearly identical results for the other parameters, with the exception of index $k$ defining the radial dependence of the effective emissivity, requiring a steeper dependence in the case of $x$-dependent than of $V_z$-dependent attenuation. This gives increased confidence in the robustness of the result. {The value, $\eta_h=0.041^{+0.028}_{-0.019}$, corresponds to a flaring angle of 5.5$^{\circ+3.8}_{-2.6}$ FWHM.}

The rotation velocity of 5.1$\pm$0.1 \kms\ {when interpreted in terms of a circular Keplerian orbit, gives} a star mass of 1.73$\pm$0.10 solar masses, to be compared with a value of $\sim$2 solar masses, typical for this type of stars \citep{Zorec2012}. We checked that the position-velocity diagrams associated with the $x$ axis ($y=0$ pixels) and $y$ axis ($x=0$ pixels) are well reproduced by the model, independently from the form chosen for the attenuation ($V_z$-dependent or $x$-dependent). However, the velocity profiles obtained on the $x$ axis near maximal intensity are significantly broader in the data than in the model, which assumes a negligibly small line width; this is further discussed in Section 4.5 below.

\subsection{Absorption from a foreground cold cloud}

The very good quality of the fits obtained in the preceding section invites further comments on the credibility of the suggested interpretation of the central attenuation as being caused by absorption from a cold cloud in the foreground.

One might indeed imagine that the disc surrounding the star is simply accreting gas from such a cloud. The absence of detectable Doppler shift of the absorption would imply that the cloud and the disc have Doppler velocities that differ by less than $\sim$1 \kms. This strongly suggests that the cloud would be closely related to 49 Ceti, probably a remnant of the original core that formed the disc several 10 Myr ago or a nearby companion core, sharing with it a common parent cloud. The age of 49 Ceti is similar to both the average formation time and average life time of molecular clouds (\citealt{Larson1994, MacLow2004}), making such an assumption plausible. CO accretion would then have to proceed toward the disc at a rate of 10$^{21}$ to 10$^{22}$ g/yr, namely $\sim$10$^{-6}$ Earth masses per year which is necessary to compensate for the loss of CO molecules (\citealt{Zuckerman2012, Moor2011}). Such an accretion rate, compared with the 2.7$\times$10$^{-4}$ Earth masses of CO gas contained in the disc, means only a third of a per cent per year, the disc taking $\sim$300 years to renew itself. It is small enough to produce no detectable effect on the gas Doppler velocity. The dynamic of the disc would then be governed by two competing factors: gas accretion proceeding from outside inward and UV dissociation proceeding from inside outward. In addition to explaining why the properties of the gas disc resemble those of a young accretion disc, such a picture would have the advantage of not having to resort to ad hoc braking and/or excitation mechanisms for the gas molecules.

Unfortunately, such a model seems to be irreconcilable with the FUV observations of \citet{Roberge2014}. Neglecting the emission of such a cloud, we can estimate which column density it takes to produce a value of $\lambda$ of 2/3, as obtained from the best fit, meaning an optical depth $\tau$ of ln$\,3$=1.1. From the relation
\begin{equation}
  N_2=\Delta v\,\tau\,\frac{g_2}{g_3}\frac{8\pi\,f^3/c^3}{A_{32}\,[1-\textrm{e}^{\frac{-(E_3-E_2)}{kT}}]}
\end{equation} 
where $\Delta v\sim\,$1.5 \kms \, is the absorption line width, $c$ is the velocity of light, $A_{32}$=2.6$\times$10$^{-6}$ s$^{-1}$ is the transition rate, \mbox{$f$=345 GHz} is the frequency, $g_\textrm{J}$=2J+1, \mbox{$E_\textrm{J}/k$ =J(J+1)$\times$2.77 K} is the energy of level J, $k$ is Boltzmann constant, $T$ is the temperature taken equal to 10 K{, a reasonable value for such a molecular cloud,} and $N_2$ is the column density for the J=2 level, we obtain $N_2$=2.1$\times$10$^{15}$ cm$^{-2}$. The CO column density $N_\textrm{CO}$ is then obtained from the relation
\begin{equation}
  N_\textrm{J}/N_\textrm{CO}=\textrm{(2J+1)}\,\textrm{exp}(-E_\textrm{J}/kT)\,2.8/T
\end{equation}
giving $N_\textrm{CO}$=8.0$\times$10$^{15}$ cm$^{-2}$ and $N_3$=0.57$\times$10$^{15}$ cm$^{-2}$

Such a CO column density is much larger than the upper limit obtained by \citet{Roberge2014} from FUV HST observations, $\sim$5$\times$10$^{12}$ cm$^{-2}$. We are unable to think of a reason that could explain such a discrepancy and we must therefore conclude that the interpretation in terms of an absorbing cloud in the foreground cannot be retained as a sensible picture of the observed attenuation.

Finally, we note that the emission per beam $F$ of such a cloud, as obtained from the relation
\begin{equation}
  F\,\Delta v =\frac{h\,c\,A_{32}\,N_3\, \Omega}{4\pi}
\end{equation}
where $h$=6.626$\times$10$^{-4}$ Jy cm$^2$ s is Planck constant and \mbox{$\Omega$=5.1$\times$10$^{-12}$ sr} is the solid angle spanned by the beam, would be $F$=80 mJy beam$^{-1}$ compared with the noise level of 8.8 mJy beam$^{-1}$. Integrating over a 12$\times$12 arcsec$^2$ square centred at the origin but excluding the region covered by $\pm$4.2 arcsec in $x$ and $\pm$1.2 arcsec in $y$ and accounting for the zero-spacing problem correction, predicts a CO flux density of \mbox{$\sim$20 Jy} {to be compared with an observed flux density not exceeding 2 Jy (3-$\sigma$ upper limit)}, again conflicting with what the model would imply.   

\subsection{Optical thickness and line width effects}
\citet{Horne1986}, and later \citet{Beckwith1993} have underlined the importance of the contribution of the velocity gradient to the shape of the global Doppler velocity profile. The point is that different Doppler velocity intervals correspond to regions of the disc having very different areas: for a disc seen edge-on, both the smaller and larger Doppler velocities are found near $\lvert x \rvert$=0 and correspond to small areas while large $\lvert x \rvert$ values relate to Doppler velocities close to the rotation velocity at the outer edge of the disc and correspond to large areas. More precisely, the outer wings of the double-horned profile correspond to small crescent-like regions near the disc centre, while the depression at small $\lvert x \rvert$ values corresponds to small-curvature arcs near the line of sight pointing to the star. The horns correspond instead to Doppler velocities of the order of the rotation velocity at the outer edge of the disc and cover a broad spiral-like sector.
\begin{figure}
\centering
\includegraphics[width=4.2cm,trim=3.cm 1.cm 2.9cm 2.4cm,clip]{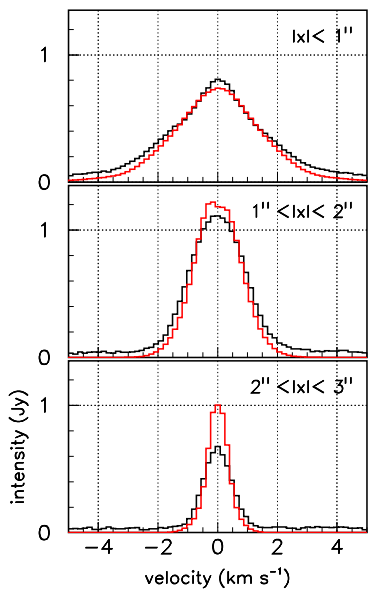}
\caption{Observed (black) and modelled (red) Doppler velocity distributions are displayed as a function of $V_z-$<$V_z$> in three intervals of $\lvert x \rvert$.}
\label{fig9}
\end{figure}

In a given pixel, the observed width of the Doppler velocity profile receives two contributions: from the line width proper and from the fact that the associated disc region spans different rotation velocities (the Keplerian shear effect). If the former is significantly larger than the latter, it smears the Doppler velocity distribution and Keplerian shear is unimportant. If instead, as in the present case, the line width is smaller than the Keplerian shear, the latter dominates.

A property of the Keplerian shear effect is to depend on optical thickness: if the disc has some significant geometrical thickness, a large optical thickness implies that emission is enhanced near the disc surface and the associated disc region is accordingly modified. As a result, {\citet{Beckwith1993}} find that the outer parts of the disc always dominate the flux density at any Doppler velocity, the decrease in brightness temperature being more than compensated for by the increase in area at large radii. This result, along with the sharply limited outer radius, causes the double-horned profile and occurs for all discs for which the temperature falls more slowly than $r^{-1}$. In particular, in the case of typical T Tauri accretion discs having masses of the order of 0.1 solar masses, {\citet{Beckwith1993}} predict a global Doppler velocity distribution increasing linearly from zero when the Doppler velocity increases from 0 to the outer disc rotation velocity. Such a profile is reminiscent of the profile observed for CO(3-2) emission from 49 Ceti and raises the question of its relevance to the present work.

However, we recall that the evidence obtained in Section 4.2 for attenuation of central velocities (or central $\lvert x \rvert$ values) rests on an argument using the intensity integrated over velocities, which, in principle, has nothing to do with Keplerian shear. Yet, the similarity between {\citet{Beckwith1993}} profiles and the 49 Ceti profile must make us cautious and beware of possible indirect effects. A second remark is that the outer edge of the gas disc of 49 Ceti is not as abrupt as assumed in the model of {\citet{Beckwith1993}}. A third remark is that the inclination of the disc differs clearly from zero, as found not only by the present analysis but also by those of \citet{Choquet2016} and \citet{Roberge2014}. A fourth remark is that the gas disc of 49 Ceti is 8 orders of magnitude less massive than the typical T Tauri disc mentioned above. Both the analyses of \citet{Hughes2008} and \citet{Roberge2013} conclude that it is optically much thinner than the discs considered by {\citet{Beckwith1993}}. A fifth remark is that the Doppler velocity profile of HD 21997 \citep{Kospal2013} does not display a narrow central dip as seen in 49 Ceti, in spite of a 200 times larger CO mass and otherwise similar geometry and temperature.

A parameter of relevance to the present discussion is the line width. Figure \ref{fig9} {compares observed line profiles with modelled profiles obtained in section 4.3 in three intervals} of $\lvert x \rvert$: $\lvert x \rvert$<1 arcsec, 1<$\lvert x \rvert$<2 arcsec and 2<$\lvert x \rvert$<3 arcsec. In each individual pixel the intensity distribution is shifted and displayed as a function of $V_z-$<$V_z$>. The width of the modelled line is exclusively due to the Keplerian shear effect when integrating over the pixel, over the beam and over the line of sight across the disc. Any difference between the observed and modelled line widths must therefore be assigned to an effective line width receiving contributions from the intrinsic line width and from whatever additional effect may be of relevance, {such as imperfections of the model}. In order of increasing values of $\lvert x \rvert$, using Gaussian fits in each interval, we find Keplerian shear line widths (obtained from the model) of respectively 3.2, 1.8 and 0.9 \kms\ FWHM. {To obtain an estimate of the maximal values of the intrinsic line widths that the data can accommodate, we subtract in quadrature these Gaussian line widths, meant to account for the Keplerian shear, from the observed line widths. We obtain this way upper limits to the intrinsic line widths of} 1.4$\pm$0.3, 1.2$\pm$0.2 and 0.8$\pm$0.1 \kms\ FWHM. Note that thermal broadening is proportional to $\sqrt{T}$ and is only 0.29 \kms\ FWHM at 50 K, while the spectral resolution is {0.1} \kms, suggesting that other effects {may contribute significant broadening}. Including in the model both thermal broadening and an ad hoc small smearing of the direction of the velocity vector would cause an increase of the effective line width particularly important in the central disc region, as is actually observed; however, we refrained from doing so as it would not add much physics insight in the present state of our understanding.

In summary, in spite of the intriguing similarity between the Doppler velocity profiles obtained by {\citet{Beckwith1993}} for T Tauri accretion discs and the 49 Ceti profile displayed in Figure \ref{fig5} (upper left panel), we do not see how the arguments developed in the former case could be of relevance to the present case.

\section{Flux density and morphology of the dust emission}

We evaluate the flux density of continuum emission at 350 GHz (0.86 mm wavelength) by summing over all pixels exceeding 3$\sigma$'s, accounting for the Gaussian beam profile  and correcting for lower contributions by extrapolating under the noise using the distribution displayed in Figure \ref{fig1}, giving 11.4$\pm$0.5 mJy, larger than the value of 8.2$\pm$1.9 mJy measured by \citet{Song2004} at 0.85 mm wavelength using the JCMT (SCUBA). The correction for the loss of short spacing cannot be reliably evaluated but is estimated, as described in Section 2, at the level of $\sim$+5 mJy. In addition, a normalization uncertainty of $\sim$10\% associated with the calibration of the absolute flux must be taken in account.

As previously noted by \citet{Hughes2008}, the above values are inconsistent with an earlier IRAM 1.2 mm measurement by \citet{Bockelee1994}. But \citet{Roberge2013} and \citet{Wahhaj2007} analyses of dust observations predict a significantly larger 0.86 mm emission of $\sim$19 mJy using a temperature of the cold component (which is of relevance here) of T=65$\pm$1 K and an emissivity power law index $\beta$=0.6$\pm$0.1. Between the low value of \citet{Song2004} and the high SED estimates, our measurement is not able to choose, the short-spacing correction being insufficiently reliable.

The effective emissivity obtained by solving integral equation (3) as done above for CO(3-2) emission and using a 1-$\sigma$ cut is displayed in Figure \ref{fig6} (right). While it might suggest the presence of separated rings of dust, the peaks nearly disappear when using a 3-$\sigma$ cut: the signal to noise ratio is insufficient to make a reliable claim for the presence of such features. Below $r$ $\sim$0.65 arcsec ($\sim$40 au) the effective emissivity is found to be zero on average. Both $x>0$ and $x<0$ observations display significant inhomogeneity, which {we ignore for the time being: in the model we simply use the third degree polynomial fit to the effective emissivity (Figure \ref{fig6} right) obtained by solving integral equation (3), thereby smoothing out the observed inhomogeneity. For $\lvert x \rvert$<0.65 arcsec ($\sim$40 au), it is set to zero.}

We recall that correcting for the zero-spacing problem would extend the radial distribution of the effective emissivity by $\sim$0.5 arcsec outward.
\begin{figure}
\centering
\includegraphics[width=7.cm,trim=0.cm 8.cm 2.5cm 0.cm,clip]{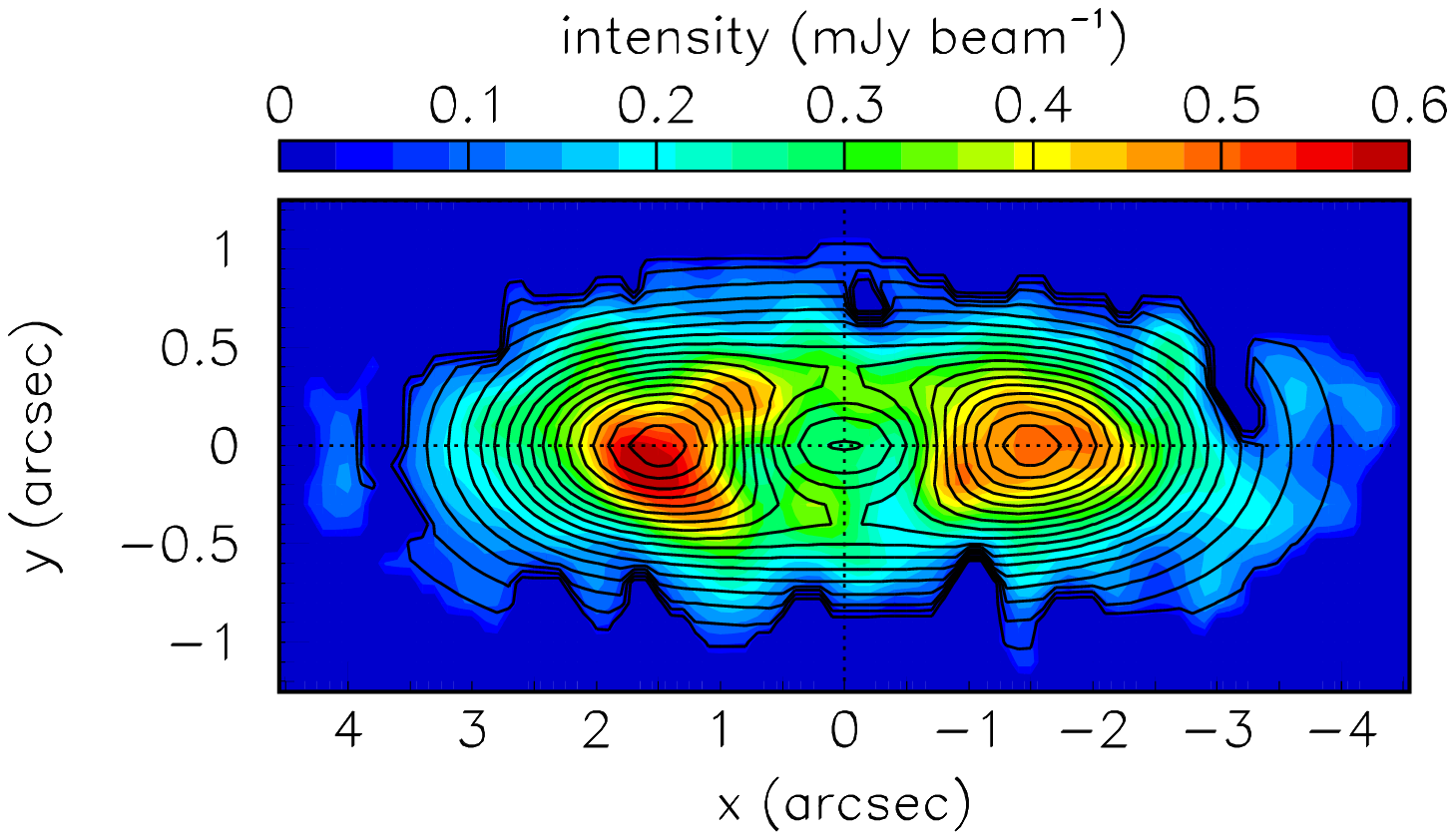}
\includegraphics[width=7.cm,trim=0.cm 8.cm 2.5cm 0.cm,clip]{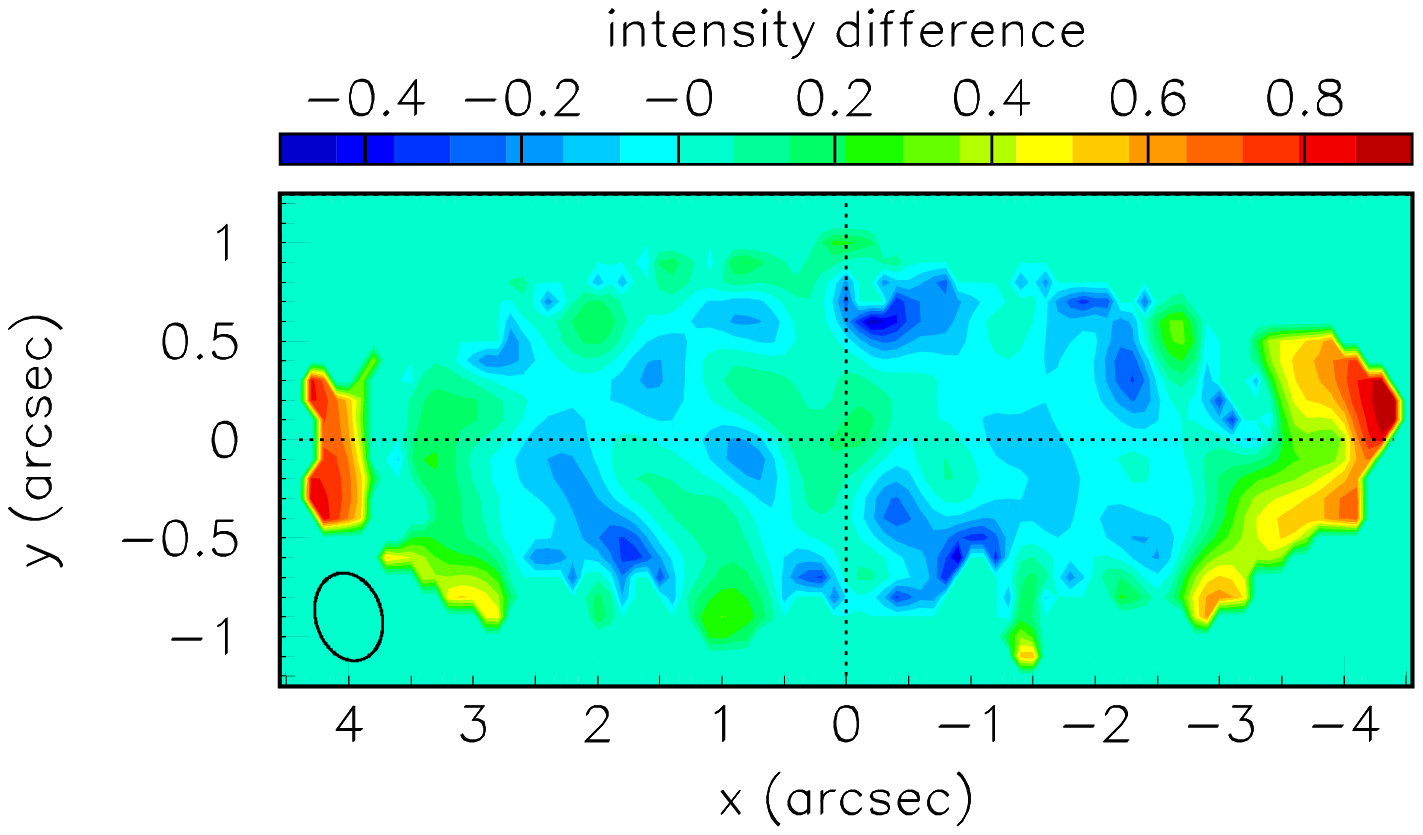}
\caption{350 GHz continuum emission. Upper panel: map of the measured (1-$\sigma$ cut, colour) and modelled (contours) intensities. Lower panel: map of the intensity difference between data and model. {Note the different $x$ and $y$ scales}}
\label{fig10}
\end{figure}

\begin{figure}
\centering
\includegraphics[height=4.5cm,trim=1.7cm 1.cm 1.5cm 2.cm,clip]{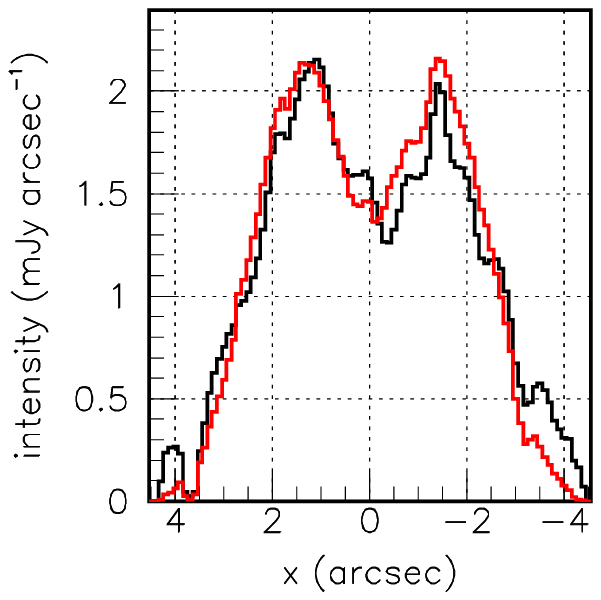}
\includegraphics[height=4.5cm,trim=2.cm 1.cm 2.3cm 2.cm,clip]{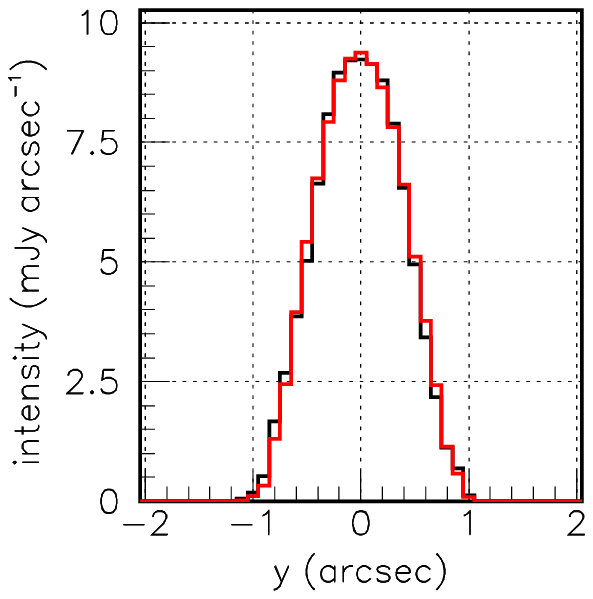}
\caption{350 GHz continuum emission. Dependence on $x$ (left) and on $y$ (right) of the continuum intensity as observed (black) and as obtained from the best fit (red).}
\label{fig11}
\end{figure}

The morphology of the disc is then described by only two parameters, the inclination $i$ with respect to the line of sight and a parameter $\Delta_h =\eta_h\,R$ describing the dependence of the effective emissivity on the distance $h$ to the disc mid-plane, which we write as a flared Gaussian as in Relation (4). The best fit gives a $\chi^2$ of 0.84 per degree of freedom and is illustrated in Figures \ref{fig10} and \ref{fig11}. The minimisation of $\chi^2$ is limited to pixels in which the measured intensity exceeds 0.055 mJy beam$^{-1}$, which explains the small irregularities observed in the model prediction. The best fit values of the parameters are $i$=11.1$^\circ\pm$1.4$^\circ$ and $\eta_h$=0.10$\pm$0.06 {corresponding to a flaring angle of 14$^\circ\pm8^\circ$ FWHM;} the gas and dust discs share a same inclination with respect to the line of sight while the latter is approximately twice as thick. However, important uncertainties are attached to the values of $\eta_h$ for both discs, the gas value covering the range between 0.02 and 0.07, the dust value between 0.04 and 0.16: the data do not strongly constrain the values adopted to describe the disc thickness. Here, as in the CO(3-2) case, uncertainties are defined by a 10\% increase of $\chi^2$, {providing a precise evaluation of the sensitivity of the fit quality to the parameter values and giving a realistic estimate of the errors attached to them. We checked that the parameters obtained for the best fit are essentially unchanged when the polynomial form used to describe the radial dependence of the effective emissivity is varied within reasonable limits.}

We note, on both Figures \ref{fig10} and \ref{fig11}, a small excess of the observed intensity at $x$-values of $\sim\pm$4 arcsec (240 au) that might indicate the presence of a dust ring surrounding the disc; however higher sensitivity observations would be necessary to confirm or infirm such an assertion. Moreover, we recall that at such large distances from the star the correction for acceptance loss due to lack of short enough baselines becomes large, corresponding to an increase of the radial extension of the disc reaching $\sim$0.5 arcsec.

\section{Summary and discussion}
\subsection{Summary}
We have presented a number of new results on the 350 GHz continuum and $^{12}$CO(3-2) emissions of the debris disc of \mbox{49 Ceti}.

1. Contrary to what had been assumed earlier, continuum and CO(3-2) emissions are not co-spatial although both share a same position angle of $\sim$17.5$^\circ$ and a same inclination angle with respect to the line of sight, $11.1^\circ\pm0.9^\circ$ for the gas and $11.1^\circ\pm1.4^\circ$ for the dust. The acceptance-corrected radial extension of the CO disc, $\sim$20 au to $\sim$210 au, is more central than that of the dust disc, $\sim$40 au to $\sim$280 au. A very recent VLT observation \citep{Choquet2016} that appeared after completion of the present work confirms these results. Both disc thicknesses are described by flared profiles having flaring angles of respectively 5.5$^{\circ+3.8}_{-2.6}$ FWHM for the gas and 14$^\circ\pm$8$^\circ$ FWHM for the dust. The CO disc is thin enough to account for the HST observations of \citet{Roberge2014}. In both cases there is room for some ambiguity between inclination angle and disc thickness.

2. CO(3-2) emission is observed to be very homogeneous and symmetric, leaving room for only small deviations, such as disc warping or density/temperature lumps, with respect to what is expected from a thin flared disc, symmetric about both the disc axis and the disc mid-plane. On the contrary, continuum emission displays spatially resolved lumps (lower panel of Figure \ref{fig10}), which we did not attempt to model.

3. Strong evidence has been obtained for a deficit of observed CO(3-2) emission in the central $x$ region, at variance with continuum emission. Its extension on the sky covers the whole $y$ range and some 50 au FWHM in $x$, smaller than the warm central emission observed in mid-infrared by \citet{Wahhaj2007}. We find it more natural to model the attenuation as resulting from absorption of the central frequencies, as would be the case if the CO disc were seen behind a cold cloud in the foreground, exactly at rest with respect to the star. The absorption covers a Doppler velocity range of 1.5$\pm$0.5 \kms\ FWHM.

4. The quality of the fits obtained with an ad hoc description of the observed attenuation mimicking absorption from a foreground cloud might suggest taking such an interpretation seriously. In that case the observed gas disc of 49 Ceti would simply be fed by accretion from such a cloud at a slow rate implying $\sim$300 yr to renew the whole gas content. The resulting accretion rate would be too small to be detectable on the Doppler velocity profile. This would explain the primordial apparent properties of the gas disc without resorting to colliding comet-like bodies. However, such a simple interpretation cannot be retained as it conflicts with the absence of CO absorption line in the HST data of \citet{Roberge2014}, the cloud CO column density being $\sim$8$\times$10$^{15}$ molecules per square centimetre, over three orders of magnitude above the detectable level. Also, no CO emission is detected from the region surrounding the disc where the cloud would be expected to contribute. 

5. We have presented models of the radial dependence of the effective emissivity of both CO(3-2) and continuum emissions. In the former case, the strong central attenuation makes it difficult to trace it reliably at small distances from the star. CO(3-2) emission is found to rise toward small radii more than continuum emission does. It is associated with the need to compensate for attenuation: a narrower attenuation region would imply a less steep radial dependence of the CO(3-2) effective emissivity.

6. Values of the intensities integrated over the source size but not corrected for the loss of short spacing have been given for both CO(3-2) emission, {\mbox{7.1$\pm$0.4 Jy \kms}}, and continuum emission, 11.4$\pm$0.5 mJy. Crude estimates of the short spacing correction have been given at the respective levels of 1.6 Jy \kms\, and 5 mJy. The gas value is in good agreement with the value measured earlier by \citet{Dent2005} but the dust value is larger than that measured by \citet{Song2004} and close to the predictions of the disc models proposed by \citet{Hughes2008} and \citet{Roberge2013}, according to which the CO mass of the disc is between $\sim$2$\times$10$^{-4}$ and $\sim$10$^{-3}$ Earth masses and the 0.86 mm SED value is $\sim$19 mJy. The crudeness of our estimated short spacing correction prevents a reliable resolution of this apparent disagreement, which will require new lower resolution observations. The use in the present work of an effective emissivity to model the detected emission mixes the effects of temperature and density and the new results do not justify repeating the detailed analysis that \citet{Hughes2008} and \citet{Roberge2013} have made of the disc structure and chemistry for both gas and dust. 

7. Kinematics has been found to be Keplerian over a very broad range of distances from the star, typically 20 au to 200 au, with a power index of $-0.46^{+0.05}_{-0.02}$ and a rotation velocity of 5.1$\pm$0.1 \kms\ at a distance of 60 au (1 arcsec) from the star, corresponding to a masss of 1.73$\pm$0.10 solar masses for the central star. The observed line width was observed to be dominated by Keplerian shear over most of the disc volume and upper limits to the intrinsic line width (FWHM) have been evaluated, decreasing from \mbox{1.4$\pm$0.3 \kms}  for \mbox{$\lvert x \rvert$<1 arcsec} to 0.8$\pm$0.1 \kms\  for 2<$\lvert x \rvert$<3 arcsec.

\subsection{Implications on the disc structure and chemistry \citep{Hughes2008}}
The very detailed analysis of the chemistry that governs the morphology and temperature of the disc that was made by \citet{Hughes2008} remains essentially valid. However, some of the assumptions made by these authors need now to be modified: the inclination angle with respect to the line of sight is definitely non zero; the dust extends to significantly larger distances from the star than the gas does, both radially and axially; the inner radius of the gas disc is nearly equal to the gravitational radius, $\sim$20 au, rather than twice this value; the lumpiness of the dust disc contrasts with the smoothness of the gas disc; the apparent warping of the gas disc suggested by the SMA data is not confirmed by the ALMA data; the central attenuation, for which the ALMA data gives strong evidence, can no longer be ignored.

Yet, the analysis made by \citet{Hughes2008} of the factors governing the physics and chemistry of the gas and dust discs remains perfectly pertinent, in particular concerning the flaring, the temperature dependence predicted by their best fit model, both across the disc and along its radius and the identification of the cooling mechanisms as due in particular to atomic species [O$_\textrm{I}$] and [C$_\textrm{II}$]. Their conclusion that the disc is optically thin remains valid, the more so now that evidence for a significant inclination with respect to the line of sight has been obtained, and we do not see how optical depth effects within the disc could cause the strong central attenuation that is observed. While the possibility for the gas to be primordial is no longer tenable with the new evaluation of the age of the disc \citep{Zuckerman2012}, the argument of \citet{Hughes2008} in favour of gas clearing proceeding from the centre out by photo-dissociation from the star remains valid. The remark that the densities in the disc are too low for efficient gas-dust coupling, implying that the gas temperature is governed by photoelectric heating and line cooling, is comforted by the observed differences between the gas and dust morphologies. The Keplerian regime observed to govern the kinematics over most of the disc volume, with no sign of inhomogeneity, presents now a stronger challenge to models that deny the primordial origin of the gas.    
\subsection{Implications on the picture of colliding comet-like objects \citep{Zuckerman2012}}
The new ALMA observations presented here provide important information that help refining and/or clarifying the model proposed by \citet{Zuckerman2012}. The model is motivated by the need to cope with the short lifetime of CO molecules against photo-dissociation, at the kyr scale, compared with the age of the disc, $\sim$40 Myr, imposing rapid and permanent production of CO, faster than dust production. The large dustiness (49 Ceti ranks among the 1-2\% dustiest of nearby A-type stars having infrared emission dominated by Kuiper Belt analogue) argues in favour of a very massive reservoir of colliding, CO-rich, comet-like bodies as the source of CO production. \citet{Zuckerman2012} demonstrate the validity of such a picture with a toy model used as illustration. Scaling from the Kuiper Belt, the toy model uses a comet reservoir shaped as a flared ring, with an inner radius of 100 au, an outer radius of 150 au, a surface density proportional to $r^{3/2}$, a mass of 400 Earth masses, a total content of 2.4$\times$10$^{14}$ comets, each 1 km in diameter and experiencing 5$\times$10$^6$ collisions per year. The model assumes that 50\% of the mass of a typical comet is water ice, that CO and CO$_2$ account for 10\% each, and that photo-dissociation of CO$_2$ yields a CO molecule. When two comets collide, 50\% of the total mass becomes debris consisting of CO, CO$_2$, and solid material composed primarily of equal {amounts} by mass of water ice and silicates. The proto-planetary disc must have had a mass of a few tenths of a solar mass and a gas to dust ratio of at least 20\% for the total mass of CO and CO$_2$ to add up to $\sim$80 Earth masses. The CO production rate is found to exceed the dust production rate by an order of magnitude, a result of the much shorter time it takes to release large amounts of CO following a disruptive impact compared to the time it takes to reduce a large chunk of solid debris down to 10 $\mu$m-size particles. With such a set of parameters, the {amount} of CO outgassed is more than enough to produce the measured 3$\times$10$^{-4}$ Earth masses.

Such a model implies therefore that the outgassed CO is a better tracer of comet collisions than the dust, which takes time to cascade down to detectable grain sizes. However, it seems difficult to reconcile this idea with the quiet and smooth Keplerian regime that governs the disc over most of its volume. Nothing in the morphology of the gas gives a hint about the location where comet collisions take place. If anything, the present observations suggest that the ring of colliding comets is probably more central than in the toy model. As remarked by \citet{Roberge2014}, this requires a braking mechanism preventing CO from escaping too quickly. Their observation of carbon overabundance in 49 Ceti induces them into suggesting a braking mechanism similar to that used by \citet{Fernandez2006} to explain the same phenomenon in $\beta$ Pictoris: the idea is that the ionized gas, in {amount} comparable with the neutral gas, couples into a single ionic fluid with an effective radiation pressure coefficient, and that carbon overabundance lowers the coefficient, keeping the whole gas disc bound to the star. The observations presented here may be at least qualitatively consistent with such a picture but the proposed braking mechanism needs to be particularly efficient to allow for such a large radial extension of the thin Keplerian gas disc.

The different morphology of the inner parts of the gas and dust discs, as observed in the present data, would suggest that most of the outgassing happens close to the star, the dust taking some time to cascade-fragment to a detectable size, at the same time as it drifts away to larger radii. Note that the angular area of the sky covered by 2$\times$10$^{14}$ opaque objects {randomly distributed}, each 1 km in diameter, is \mbox{$\sim$0.01 au$^2$}, too small to cause a significant obscuration; when fragmented into 1 m diameter opaque objects, they cover \mbox{10 au$^2$}, but they would obscure preferentially half of the central part of the disc, the part which is in their background; the part in the foreground would be less affected, contrary to what is observed.

The above model brings another new question: How are the CO molecules excited if the amount of H$_2$ is orders of magnitude lower than in normal ISM? \citet{Zuckerman2012} suggest that electrons, freed from the ionization of the copious carbon atoms, might be the answer. High resolution detection of atomic species over the whole disc would be necessary to answer this new question.

What happens near the star remains an open question: one needs to understand what relates the belt of colliding comets and the warm, fine-grained, atomic-species-rich dust component observed in mid-infrared. 
\subsection{Comparison with other debris discs}
A recent review of so-called hybrid debris discs, containing both gas and dust, has been recently compiled by \citet{Pericaud2016}. As was mentioned in the introduction, CO emission has been detected from four other debris discs, $\beta$ Pictoris (\citealt{Fernandez2006, Dent2014}), HD 21997 (\citealt{Moor2011, Moor2013, Kospal2013}), \mbox{HD 141569} \citep{White2016} and HD 131835 \citep{Moor2015}. Table \ref{Table 2} lists a few relevant parameters.

\begin{table*}
\centering 
\caption{Debris discs with CO emission. Numbers have been rounded and/or averaged; references should be consulted to obtain actual values. {The dust mass is estimated over the whole FIR range.}}
\label{Table 2}
\begin{tabular}{|c|c|c|c|c|c|c|}
\hline 
\multirow{1}{*}{Star name} & \makecell{Age \\(Myr)} & \makecell{Dust mass \\(10$^{-3}$ M$_\textrm{Earth}$)} & \makecell{CO mass  \\(10$^{-3}$ M$_\textrm{Earth}$)} & \makecell{R$_\textrm{gas}$ in/out \\(au)} & \makecell{R$_\textrm{dust}$ in/out \\(au)} & \makecell{T$_\textrm{dust}$ \\(K)} \\
\hline 
49 Ceti & 40 & 300 & 0.3 & 20/{210} & 40/280 & 70/170 \\
\hline
HD 21997 & 30 & 90 & 60 & 30/140 & 60/150 & 60 \\
\hline
$\beta$ Pictoris & 16 & 80 & 0.03 & 50/160 & $-$ & {80-100} \\
\hline
HD 131835 & 16 & 500 & 0.5 & {35/120} & {35/310} & 70/180 \\
\hline
HD 141569 & 5 & 40 & 2 & 30/210 & {10/210} & $-$ \\
\hline
\end{tabular} 
\end{table*}
The morphology of the gas discs, and of the dust discs when available, are remarkably similar. In both the 49 Ceti and HD 21997 cases, where both gas and dust discs have been observed with high spatial resolution, the inner radius of the gas disc is significantly smaller than that of the dust disc.

While dust masses span only one order of magnitude, gas masses span four orders of magnitudes. HD 21997 is the disc that contains most gas, with a gas to dust ratio approaching unity, followed by HD 141569, with a gas to dust ratio of 5\%. HD 131835 and 49 Ceti have gas to dust ratios at per mil level and $\beta$ Pictoris below 4$\times$10$^{-4}$. When taking the age in consideration, 49 Ceti and HD 21997 stand out as having much larger relative {amounts} of gas. For the three other discs, which are younger, the primordial gas scenario is more tenable. However, if one accepts the interpretation given by \citet{Moor2013} and \citet{Kospal2013} of the HD 21997 disc being made of a mixture of primordial gas and secondary dust, there is no reason not to accept it as well for 49 Ceti, which is only 10 Myr older and contains only three times as much dust but 300 times less gas.

The disc of $\beta$ Pictoris shares with that of 49 Ceti the peculiarity of having a strong carbon excess but its morphology is so different from that of 49 Ceti, for both gas and dust, and its gas to dust ratio is so much smaller in spite of its younger age, that it seems difficult to pursue the comparison farther. Indeed, as argued by \citet{Dent2014}, the disc of $\beta$ Pictoris is dominated by the presence of a gas clump, which they interpret as a region of enhanced collision either from a mean motion resonance with an unseen giant planet or from the remnants of a collision of Mars-mass planets. Nothing of that kind is present in the disc of 49 Ceti.

Finally, 49 Ceti is unique in displaying a strong attenuation in the centre of the Doppler velocity spectrum, clearly absent from HD 21997 and HD 141569, the only two other discs for which accurate spectra are available.

{One cannot be satisfied with unrelated ad hoc explanations meant to solve each individual puzzle separately; one needs instead a picture that explains the whole set of observations as simply and coherently as possible \citep{Pericaud2016}. New probes should be used to explore the discs, including their central region, with high spatial and spectral resolutions; in particular, ALMA observations of atomic species and of CO emission from other excitations than CO(3-2), such as CO(1-0) and CO(7-6), would be an important asset. In addition, one needs a detailed understanding of the evolution of the molecular cloud from which 49 Ceti was formed.}

\subsection{Conclusion}
In summary, we have presented a rich set of new observations making use of the high sensitivity and excellent spectral and spatial resolutions of ALMA, which contribute significant addition to our knowledge of the debris disc of \mbox{49 Ceti}. We have found evidence for a strong central attenuation of CO(3-2) emission, which we have shown to be well described by absorption from a cold gas cloud in the foreground. However, such a description conflicts with the FUV observations of \citet{Roberge2014} and cannot be retained. The evidence for a deficit of the observed \mbox{CO(3-2)} emission for Doppler velocities approaching the systemic velocity is strong and calls for a plausible mechanism that could cause it.

We noted an intriguing similarity between the Doppler velocity profile of CO(3-2) emission and those studied by \citet{Beckwith1993} for the accretion discs of T Tauri stars, described as the combined effect of Keplerian shear and optical thickness; however, we were unable to extend such a description to the case of 49 Ceti in a sensible way; yet, we cannot exclude that some combination of geometry, temperature and density, with important optical thickness effects, could explain the effect; we were simply unable to think of one that could make sense.

\section*{Acknowledgements}
We thank the anonymous referee for very valuable comments that helped improving greatly the quality of the manuscript. This paper makes use of the following ALMA data: ADS/JAO.ALMA\#2012.1.00195.S. ALMA is a partnership of ESO (representing its member states), NSF (USA) and NINS (Japan), together with NRC (Canada), NSC and ASIAA (Taiwan), and KASI (Republic of Korea), in cooperation with the Republic of Chile. The Joint ALMA Observatory is operated by ESO, AUI/NRAO and NAOJ. The data are retrieved from the JVO portal (http://jvo.nao.ac.jp/portal) operated by the NAOJ. We are indebted and very grateful to the ALMA partnership, who are making their data available to the public after a one year period of exclusive property, an initiative that means invaluable support and encouragement for Vietnamese astrophysics. We particularly acknowledge friendly support from the staff of the ALMA Helpdesk. We thank Dr. Anne Dutrey, Dr. St\'ephane Guilloteau and Dr. Pierre Lesaffre for a critical reading of the manuscript. We are particularly grateful to Dr. Anne Dutrey and Dr. St\'ephane Guilloteau for useful suggestions that helped making the description of the central attenuation problem clearer.  We thank the scientific editor for useful comments and advice. Financial support is acknowledged from the Vietnam National Satellite Centre (VNSC/VAST), the World Laboratory, the Odon Vallet Foundation and the Rencontres du Viet Nam. This research is funded by Vietnam National Foundation for Science and Technology Development (NAFOSTED) under grant number 103.99-2016.50

After having submitted the manuscript for publication and having sent a copy to the PI, Pr M. Hughes, she sent us a copy of the work of her team on the same data \citep{Hughes2017}, more concerned with the temperature and density distribution and chemical composition, and commented on the complementarity of the two studies, for which we wish to express our deep gratitude.








\appendix

\bsp	
\label{lastpage}
\end{document}